\documentclass[showpacs]{revtex4} 
\usepackage{epsfig,amsmath,amssymb,graphicx,upgreek,textcomp}
\usepackage{natbib}
\usepackage[english]{babel} %

\begin{document}

\vspace{0mm}
\title{Does a massless Goldstone boson exist?} %
\author{Yu.M. Poluektov}
\email{yuripoluektov@kipt.kharkov.ua (y.poluekt52@gmail.com)} %
\affiliation{National Science Center ``Kharkov Institute of Physics and Technology'', 61108 Kharkov, Ukraine} %

\begin{abstract}
Classical and quantum complex nonlinear scalar fields are
considered. A new approach to the quantization of nonlinear fields
and the construction of a perturbation theory with allowance for
spontaneous symmetry breaking is proposed, based on the use of the
relativistic model of a self-consistent field as the main
approximation. The concept of a particle is analyzed within the
framework of the theory of nonlinear quantum fields. When
constructing single-particle states, the contribution of vacuum
fluctuations is systematically taken into account.  Within the
framework of the developed approach, the problem of the existence of
massless scalar particles is discussed. It is shown that successive
consideration of the vacuum averages of not only one field operator,
but also the products of two field operators, leads to the
appearance of masses for scalar particles. Various states in which
the field can exist for given parameters entering into the
Lagrangian are considered, and the vacuum energy densities in these
states are calculated. It is shown that, depending on the values of
the parameters entering into the Lagrangian, the vacuum energy
density can be either positive or negative, which is important for
modern cosmology.
\newline%
\end{abstract}
\maketitle

\section{Introduction}\vspace{-0mm} 
The study of states with broken symmetries in quantum field theory
began under the influence of works on the microscopic theory of
superfluidity [1] and superconductivity [2--4] in works [5--7]. The
effect of spontaneous gauge symmetry breaking is an important
element of the modern theory of elementary particles [8--13]. Most
often it is implemented by including massive scalar fields with the
``wrong'' sign of the mass square into the Lagrange function (the
Goldstone mechanism [7]). This mechanism was used in the
construction of a unified theory of weak and electromagnetic
interactions [9].

Goldstone [7] considered a nonlinear complex scalar field, in the
Lagrangian of which the parameter that determines the square of the
particle mass in the absence of nonlinearity can be negative. For
the nonrelativistic case, a similar problem in the phenomenological
theory of superconductivity was previously studied by Ginzburg and
Landau [2]. Goldstone showed that the field considered by him
describes two types of scalar particles, one of which has a finite
mass, and the other particle is massless (the Goldstone boson). In
work [8] the relationship between the gauge symmetry breaking and
the existence of a massless boson was shown.

Scalar particles play an important role in the modern theory of
elementary particles. L.B.\,Okun called the area where theorists
study scalar particles ``Scalarland'' [13]. In recent years,
experiments on detection of the scalar Higgs boson have caused great
resonance. The development of methods for describing quantum
relativistic scalar fields, especially nonlinear fields in states
with broken symmetries, continues to be an urgent task for both
physics and cosmology [14].

The aim of this work is to analyze the methods of classical and
quantum description of a neutral complex nonlinear scalar field,
with taking into account the possible breaking of the phase (or
gauge) symmetry, and to develop a new approach to constructing a
perturbation theory. The problem of defining the concept of a
particle in the framework of the nonlinear field theory for cases
where the nonlinearity is not assumed to be small is considered. On
an example of a complex nonlinear scalar field, we propose a new
unified approach to the description of the states of nonlinear
relativistic quantum fields with spontaneously broken symmetries,
which consistently takes into account the role of vacuum
fluctuations. In the developed approach the complete Lagrangian of a
nonlinear field, as it is usually accepted, is decomposed into two
parts, one of which describes free particles and the other describes
their interaction. The novelty of the method lies in the choice of
the leading approximation and the way of constructing the
approximating Lagrange function of noninteracting particles.

The Lagrange function of free particles is chosen as a quadratic
form with respect to field operators, which also contains terms with
a lower symmetry than the symmetry of the original Lagrangian. The
parameters entering into the Lagrangian of free particles are found
from the requirement of its maximum proximity to the exact Lagrange
function. The system of nonlinear equations obtained on the basis of
such a condition makes it possible to determine the parameters of
the free Lagrangian, in particular, the masses of the field
particles. These equations can have different solutions
corresponding to states with different symmetries. The proposed
approach allows to take into account the contribution of certain
nonlinear effects already at the stage of constructing
single-particle states. In particular, the value of the interaction
constant in the initial Lagrangian determines the masses of
particles, which turn out to be different in different states. The
perturbation theory constructed on the basis of such a choice of the
main approximation proves to be applicable also in the case when the
interaction constant is not small.  It is important to note that,
with the postulated method of constructing single-particle states,
the interaction Hamiltonian without additional assumptions takes the
form of a normally ordered product of field operators.

The method formulated in this paper for describing spontaneous
symmetry breaking in quantum field theory generalizes to the
relativistic case the self-consistent quantum field approach to the
microscopic description of nonrelativistic many-particle Fermi and
Bose systems with broken symmetries, developed earlier in works
[15--18].

Within the framework of the proposed quantum field approach, not
only the vacuum averages of one field operator violating the phase
symmetry, but also the vacuum averages of the products of two field
operators are consistently taken into account, which results in the
appearance of mass for scalar particles. Due to accounting for the
vacuum averages of pairs of operators, there are no massless
Goldstone bosons in the theory. It is also shown that the particles'
vacuum energy density depends on the state in which the field stays,
and can be both positive and negative, which is fundamental for
modern cosmology.

\section{Classical complex nonlinear scalar field}\vspace{-0mm} %
Let us consider first a classical complex scalar field
$\phi\equiv\phi(x)$, depending on spatial coordinates ${\bf x}$ and
time $t$, so that $x\equiv x_\mu=({\bf x},it)$, $\mu=1,2,3,4$. This
field is characterized by the density of the Lagrange function
\begin{equation} \label{01}
\begin{array}{l}
\displaystyle{%
   \Lambda=-\frac{\partial\phi^*}{\partial x_\mu}\frac{\partial\phi}{\partial x_\mu} %
   -\kappa^2\phi^*\phi - \frac{g}{4!}\big(\phi^{*2}\phi^2\big). %
}
\end{array}
\end{equation}
We use the metric $ab={\bf a}{\bf b}+a_4b_4={\bf a}{\bf b}-a_0b_0$.
It is assumed that the sign of the real parameter $\kappa^2$, which
we will call the mass parameter, can be arbitrary, and the
interaction constant $g$ is always positive. Here and in what
follows we use the system of units $\hbar=c=1$. The field with
Lagrangian (1) obeys the nonlinear equation
\begin{equation} \label{02}
\begin{array}{l}
\displaystyle{%
  \Delta\phi - \ddot{\phi} - \kappa^2\phi - \frac{g}{12}|\phi|^2\phi=0. %
}%
\end{array}
\end{equation}
The dots at the top denote differentiation with respect to time $t$.
The energy-momentum tensor of the field has the form
\begin{equation} \label{03}
\begin{array}{l}
\displaystyle{%
   T_{\nu\mu}=-\frac{\partial\phi^*}{\partial x_\mu}\frac{\partial\phi}{\partial x_\nu} %
              -\frac{\partial\phi^*}{\partial x_\nu}\frac{\partial\phi}{\partial x_\mu} %
              -\delta_{\nu\mu}\Lambda. %
}%
\end{array}
\end{equation}
The law of conservation of energy-momentum is fulfilled
\begin{equation} \label{04}
\begin{array}{l}
\displaystyle{%
   \frac{\partial T_{\nu\mu}}{\partial x_\mu}=0. %
}%
\end{array}
\end{equation}
The energy density of the complex scalar field is given by the expression %
\begin{equation} \label{05}
\begin{array}{l}
\displaystyle{%
  {\rm H}=T_{44}=\dot{\phi}^*\dot{\phi}+\nabla\phi^*\nabla\phi+\kappa^2\phi^*\phi+g_0\phi^{*2}\phi^2.  %
}%
\end{array}
\end{equation}
Here and below, along with the constant $g$, we also use the
notation $g_0\equiv g\big/4!$.

Consider first a spatially homogeneous, stationary solution of
Eq.\,(2). In this case
\begin{equation} \label{06}
\begin{array}{l}
\displaystyle{%
  \phi_0\bigg(\kappa^2+\frac{g}{12}\,\big|\phi_0\big|^2\bigg)=0.  %
}%
\end{array}
\end{equation}
If the mass parameter is positive $\kappa^2>0$, the expression in
brackets cannot vanish, and equation (6) has a unique solution $\phi_0=0$. %
If the mass parameter is negative $\kappa^2<0$, then in addition to
the solution $\phi_0=0$ there is also a solution $\phi_0\neq 0$
whose modulus is determined by the expression
\begin{equation} \label{07}
\begin{array}{l}
\displaystyle{%
  \big|\phi_0\big|^2=-\frac{12\kappa^2}{g},  %
}%
\end{array}
\end{equation}
and the phase can be chosen arbitrarily. Since, according to (5),
the energy density of the equilibrium state is determined by the formula %
\begin{equation} \label{08}
\begin{array}{l}
\displaystyle{%
  \varepsilon_V=\kappa^2\big|\phi_0\big|^2 + \frac{g}{4!}\big|\phi_0\big|^4,  %
}%
\end{array}
\end{equation}
then in the state (7) the energy density is equal to
\begin{equation} \label{09}
\begin{array}{l}
\displaystyle{%
  \varepsilon_V=-\frac{6\kappa^4}{g}  %
}%
\end{array}
\end{equation}
and it proves to be lower than in the state with $\phi_0=0$, where $\varepsilon_V=0$. %
Thus, at $\kappa^2<0$ the equilibrium scalar field is nonzero, and
its modulus is determined by formula (7). In this case, there is a
phase degeneracy.

Let us consider small oscillations of the field relative to the
equilibrium state. At $\kappa^2>0$, for small oscillations of the
field $\phi\approx\phi'$ in (2) only the terms linear in the field
should be left; as a result, we arrive at the Klein-Gordon-Fock (KGF) equation %
\begin{equation} \label{10}
\begin{array}{l}
\displaystyle{%
  \Delta\phi' - \ddot{\phi'} - \kappa^2\phi' = 0. %
}%
\end{array}
\end{equation}
In the case $\kappa^2<0$ the scalar field can be represented as
$\phi\approx\phi_0+\phi'$, so that the linearization of Eq.\,(2) gives %
\begin{equation} \label{11}
\begin{array}{l}
\displaystyle{%
  \Delta\phi' - \ddot{\phi'} + \kappa^2\phi' - \frac{g}{12}\phi_0^2\phi'^* = 0. %
}%
\end{array}
\end{equation}
In the following, it is convenient to separate the real and
imaginary parts in the complex scalar function
\begin{equation} \label{12}
\begin{array}{l}
\displaystyle{%
  \phi=\frac{1}{\sqrt{2}}\big(\phi_1+i\phi_2\big), \qquad  \phi^*=\frac{1}{\sqrt{2}}\big(\phi_1-i\phi_2\big). %
}%
\end{array}
\end{equation}
The equilibrium condition (7) in terms of the real and imaginary
parts takes the form
\begin{equation} \label{13}
\begin{array}{l}
\displaystyle{%
  \phi_{10}^2 + \phi_{20}^2 = -\frac{24}{g}\kappa^2. %
}%
\end{array}
\end{equation}
This relation will be always satisfied if the squares of the real
and imaginary parts are presented in the form
\begin{equation} \label{14}
\begin{array}{l}
\displaystyle{%
  \phi_{10}^2 = -\frac{24}{g}\kappa^2\cos^2\alpha, \qquad \phi_{20}^2 = -\frac{24}{g}\kappa^2\sin^2\alpha, %
}%
\end{array}
\end{equation}
where $\alpha$ is an arbitrary phase angle over which, as was noted,
there is a degeneracy. The linearized equation (11) leads to two
equations for the functions
\begin{equation} \label{15}
\begin{array}{l}
\displaystyle{%
  \theta\equiv\phi_{10}\phi_1' + \phi_{20}\phi_2',  %
}%
\end{array}
\end{equation}
\vspace{-5mm}%
\begin{equation} \label{16}
\begin{array}{l}
\displaystyle{%
  \xi\equiv\phi_{20}\phi_1' - \phi_{10}\phi_2',  %
}%
\end{array}
\end{equation}
namely
\begin{equation} \label{17}
\begin{array}{l}
\displaystyle{%
  \Delta\theta - \ddot{\theta} - 2\big|\kappa^2\big|\,\theta = 0,  %
}%
\end{array}
\end{equation}
\vspace{-5mm}%
\begin{equation} \label{18}
\begin{array}{l}
\displaystyle{%
  \Delta\xi - \ddot{\xi} = 0.  %
}%
\end{array}
\end{equation}
The first equation is the KGF equation for scalar particles with the
mass $\sqrt{2}\,|\kappa|$, and the second is the wave equation for
massless particles. Choosing in (14) the phase $\alpha=0$, we obtain %
\begin{equation} \label{19}
\begin{array}{l}
\displaystyle{%
  \phi_{10}^2 = -\frac{24}{g}\kappa^2, \qquad \phi_{20}^2 = 0, %
}%
\end{array}
\end{equation}
so that $\theta\equiv|\phi_{10}|\phi_1'$, $\xi\equiv -|\phi_{10}|\phi_2'$, %
and equation (17) describes oscillations of the real part, and
equation (18) of the imaginary part of the complex field. Thus, as
was first shown by Goldstone [7], the classical nonlinear complex
scalar field with a negative mass parameter $\kappa^2<0$ has two
types of weak excitations. One type of excitations is similar to
optical oscillations with an energy gap in condensed media, and the
second type is similar to gapless sound oscillations. In the
language of particles (and in condensed matter -- quasi-particles!)
this means that there exist particles with a finite rest mass, and
also massless particles (Goldstone bosons).

It is instructive to obtain the same results by separating the
modulus and phase in the complex field $\phi=\eta e^{i\chi}$. Then
the equations of motion take the form
\begin{equation} \label{20}
\begin{array}{l}
\displaystyle{%
  \Delta\eta - \ddot{\eta} - \eta\Big[ \big(\nabla\chi\big)^2 - \dot{\chi}^2 \Big] - \kappa^2\eta - \frac{g}{12}\,\eta^3 = 0,  %
}%
\end{array}
\end{equation}
\vspace{-5mm}%
\begin{equation} \label{21}
\begin{array}{l}
\displaystyle{%
  \eta\Delta\chi - \eta\ddot{\chi} + 2\nabla\eta\nabla\chi - 2\,\dot{\eta}\,\dot{\chi} =0. %
}%
\end{array}
\end{equation}
The linearization of these equations for $\kappa^2>0$ gives the
usual KGF equation
\begin{equation} \label{22}
\begin{array}{l}
\displaystyle{%
  \Delta\eta' - \ddot{\eta'} - \kappa^2\eta' = 0. %
}%
\end{array}
\end{equation}
For $\kappa^2<0$ the modulus oscillations are also determined by the
KGF equation for the mass $\sqrt{2}\,|\kappa|$, and the phase
oscillations are determined by the wave equation
\begin{equation} \label{23}
\begin{array}{l}
\displaystyle{%
  \Delta\chi - \ddot{\chi} = 0, %
}%
\end{array}
\end{equation}
describing massless particles. Thus, Goldstone sound excitations are
associated with the phase oscillations of the complex scalar field.

If we choose the phase according to (19) and write the real and
imaginary parts as $\phi_1=\phi_{10}+\phi_1'$ and $\phi_2=\phi_2'$,
then the Lagrange function (1) can be represented as a sum
\begin{equation} \label{24}
\begin{array}{l}
\displaystyle{%
  \Lambda=\frac{6\kappa^4}{g} + \Lambda_0+\Lambda_3+\Lambda_4,  %
}%
\end{array}
\end{equation}
where
\begin{equation} \label{25}
\begin{array}{l}
\displaystyle{%
  \Lambda_0=-\frac{1}{2}\bigg[ \left(\frac{\partial\phi_1'}{\partial x_\mu}\right)^{\!2} + \left(\frac{\partial\phi_2'}{\partial x_\mu}\right)^{\!2}   \bigg]   %
  - |\kappa|^2 \phi_1'^2,
}%
\end{array}
\end{equation}
\vspace{-3mm}%
\begin{equation} \label{26}
\begin{array}{l}
\displaystyle{%
  \Lambda_3=-g_0\phi_{10}\big(\phi_1'^2+\phi_2'^2\big)\phi_1',   %
}%
\end{array}
\end{equation}
\vspace{-3mm}%
\begin{equation} \label{27}
\begin{array}{l}
\displaystyle{%
  \Lambda_4=-\frac{g_0}{4}\big(\phi_1'^2+\phi_2'^2\big)^2.   %
}%
\end{array}
\end{equation}
Lagrangian (25) describes two scalar real fields, one of which has
the mass $\sqrt{2}\,|\kappa|$ and the second has zero mass, and
Lagrangians (26) and (27) describe the interaction of these fields.
Note that we could operate not fixing the phase in (14), but work in
terms of the fields $\theta$ (15) and $\xi$ (16).

\section{Quantum complex nonlinear scalar field}\vspace{-0mm} %
When constructing a quantum field theory, it is customary to
consider classical fields at the initial stage, as was done in the
previous section, and at the subsequent stage one carries out
``quantization'' of classical fields by passing to secondarily
quantized operators obeying certain commutation relations. Such an
approach is undoubtedly justified both from historical and
methodological perspectives. However, due to the fact that the
quantum description is a deeper level of the description of reality,
it seems more natural to proceed directly from the quantum
consideration. Classical fields in such an approach should be the
limiting case of quantum fields.

In the developed here method of quantum description, which for a
real scalar field was proposed in [19] and for interacting Fermi and
Bose fields in [20], the full Lagrangian of a nonlinear field is
divided, as usually, into the sum of two parts, one of which
describes free particles and the other their interaction. The
novelty of the proposed approach lies in the method of constructing
the Lagrangian of noninteracting particles. It is assumed that the
Lagrange function of free particles contains terms no higher than
quadratic in the field operators, including those whose symmetry is
lower than the symmetry of the original Lagrangian. The parameters
of the free Lagrangian are selected from the requirement of its
closest proximity to the exact Lagrange function. The system of
nonlinear equations obtained from this condition determines the
parameters of the free Lagrangian, including the particle masses.
These equations can have several solutions, which can describe the
field states with different symmetries.   The proposed approach
allows to take into account the nonlinearity of the field already at
the stage of constructing single-particle states in a
self-consistent approximation. The perturbation theory based on such
a choice of the main approximation can also be applied in the case
when the interaction constant is not small.

Usually, the Lagrange function of the quantum field is chosen in the
normally ordered form [21,22]. This choice means that from the very
start one ignores the effects associated with zero oscillations and
playing an important role in quantum mechanics. As is known, the
vacuum fluctuations of an electromagnetic field lead to observable
physical effects, such as the Lamb shift and the Casimir effect, and
therefore should be taken into account in a consistent quantum
theory of any field. In the case of condensed matter, the role of
zero fluctuations manifests itself, for example, in the fact that at
low pressures helium remains liquid down to the lowest temperatures.
The consistent account for the vacuum fluctuations is all the more
important in the case of states with broken symmetry, which can be
characterized by the nonzero vacuum averages of the field operators.

It is easy to verify the fundamental importance of accounting for
fluctuations using the elementary example of a quantum oscillator
with the Hamiltonian $H=\big(\varepsilon/2\big)\big(a^+a+aa^+\big)=\varepsilon\big(a^+a+1/2\big)$, %
$a, \, a^+\!$ being the annihilation and creation operators. The
application of the normal ordering operation leads to the
Hamiltonian $H=\varepsilon\,a^+a$, in which the contribution of zero
oscillations is absent. The idea of constructing a perturbation
theory for a nonlinear complex scalar field implemented in this
article is illustrated in works [23,24] using a simple example of an
anharmonic oscillator. Thus, if we proceed from the Lagrangian in
which operators are written in a normally ordered form, we
immediately exclude the consideration of field's zero oscillations.
In the method of constructing one-particle states postulated in the
developed approach, the initial Hamiltonian is not assumed to be
normally ordered. Thus, from the very beginning the effects caused
by zero fluctuations are included in the theory.  It is also
important that the interaction Hamiltonian, without any additional
assumptions, takes the form of a normally ordered product of the
field operators. This has the effect that a large number of diagrams
are excluded from the series of a perturbation theory, namely all
diagrams containing loops. As a result, the structure of a
perturbation theory is substantially simplified.

Let us proceed to consideration of a quantum relativistic Bose
system, which is described by a field second-quantized operator
$\phi\equiv\phi(x)$, depending on spatial coordinates ${\bf x}$ and
time $t$, where, as before, $x\equiv x_\mu=({\bf x},it)$,
$\mu=1,2,3,4$. Such a system is characterized by the density of the
Lagrange function
\begin{equation} \label{28}
\begin{array}{l}
\displaystyle{%
   \Lambda=-\frac{\partial\phi^+}{\partial x_\mu}\frac{\partial\phi}{\partial x_\mu} %
   -\kappa^2\phi^+\phi - \frac{g}{4!}\big(\phi^{+2}\phi^2\big). %
}
\end{array}
\end{equation}
As in the previous section, it is assumed that the sign of the real
mass parameter $\kappa^2$ can be arbitrary, and the interaction
constant is positive.  The field obeys the nonlinear operator equation %
\begin{equation} \label{29}
\begin{array}{l}
\displaystyle{%
   \frac{\partial^2\phi}{\partial x_\mu^2}-\kappa^2\phi - \frac{g}{12}\phi^+\phi^2=0.  %
}
\end{array}
\end{equation}
The energy-momentum tensor of the field has the form
\begin{equation} \label{29}
\begin{array}{l}
\displaystyle{%
   T_{\nu\mu}=\frac{\partial\Lambda}{\partial\big(\partial\phi\big/\partial x_\mu\big)}\frac{\partial\phi}{\partial x_\nu} +  %
              \frac{\partial\phi^+}{\partial x_\nu}\frac{\partial\Lambda}{\partial\big(\partial\phi^+\big/\partial x_\mu\big)} - \delta_{\nu\mu}\Lambda =  %
               -\frac{\partial\phi^+}{\partial x_\mu}\frac{\partial\phi}{\partial x_\nu} -\frac{\partial\phi^+}{\partial x_\nu}\frac{\partial\phi}{\partial x_\mu} - \delta_{\nu\mu}\Lambda, %
}
\end{array}
\end{equation}
so that
\begin{equation} \label{31}
\begin{array}{l}
\displaystyle{%
   \frac{\partial T_{\nu\mu}}{\partial x_\mu}=0,  %
}
\end{array}
\end{equation}
and the 4-vector of energy-momentum is
\begin{equation} \label{32}
\begin{array}{l}
\displaystyle{%
   P_\mu=i\int\! d{\bf x}\, T_{\mu4}.   %
}
\end{array}
\end{equation}
In addition to the complex field
\begin{equation} \label{33}
\begin{array}{l}
\displaystyle{%
  \phi=\frac{1}{\sqrt{2}}\big(\phi_1+i\phi_2\big),  %
}%
\end{array}
\end{equation}
we also define the canonical momenta %
$\displaystyle{\pi\equiv\frac{\partial\Lambda}{\partial\dot{\phi}}}=\dot{\phi}^+
= \frac{1}{\sqrt{2}}\big(\dot{\phi_1}-i\dot{\phi_2}\big)= \frac{1}{\sqrt{2}}\big(\pi_1+i\pi_2\big)$. %
A complex field with coordinate and momentum $\big(\phi,\pi\equiv\dot{\phi}^+\big)$ %
is equivalent to two real fields $\big(\phi_1,\pi_1\equiv\dot{\phi_1}\big)$, $\big(\phi_2,\pi_2\equiv -\dot{\phi_2}\big)$. %
The density of the Hamiltonian of the system is given by
\begin{equation} \label{34}
\begin{array}{l}
\displaystyle{%
   {\rm H}=T_{44}=\pi\pi^+ + \nabla\phi^+\nabla\phi + \kappa^2\phi^+\phi + g_0\phi^{+2}\phi^2,   %
}
\end{array}
\end{equation}
where, as before, $g_0\equiv g\big/4!$. The energy flux density,
equal to the momentum density, has the form
\begin{equation} \label{35}
\begin{array}{l}
\displaystyle{%
   p_i=iT_{4i}=-\nabla_i\phi^+\dot{\phi} - \dot{\phi}^+\nabla_i\phi,  %
}
\end{array}
\end{equation}
and the momentum flux density is defined by the formula
\begin{equation} \label{36}
\begin{array}{l}
\displaystyle{%
   \sigma_{ik}=-T_{ik}=\nabla_i\phi^+\nabla_k\phi + \nabla_k\phi^+\nabla_i\phi.  %
}
\end{array}
\end{equation}

For coinciding times we postulate the standard commutation relations
\begin{equation} \label{37}
\begin{array}{l}
\displaystyle{%
   \big[\pi({\bf x},t),\phi({\bf x}',t)\big]=-i\delta({\bf x}-{\bf x}'), \qquad   %
   \big[\phi({\bf x},t),\phi({\bf x}',t)\big]=\big[\pi({\bf x},t),\pi({\bf x}',t)\big]=0.   %
}
\end{array}
\end{equation}
The field operators in (28) and subsequent formulas are taken in the
Heisenberg representation:
\begin{equation} \label{38}
\begin{array}{l}
\displaystyle{%
   \phi({\bf x},t)=e^{iHt}\phi({\bf x},0)e^{-iHt}, \qquad  \pi({\bf x},t)=e^{iHt}\pi({\bf x},0)e^{-iHt},  %
}
\end{array}
\end{equation}
where $\phi({\bf x},0),\, \pi({\bf x},0)$ are operators in the
Schr\"{o}dinger representation. For real fields the commutation relations hold %
\begin{equation} \label{39}
\begin{array}{l}
\displaystyle{%
   \big[\pi_1({\bf x},t),\phi_1({\bf x}',t)\big]=-i\delta({\bf x}-{\bf x}'), \qquad   %
   \big[\pi_2({\bf x},t),\phi_2({\bf x}',t)\big]=i\delta({\bf x}-{\bf x}'),    %
}
\end{array}
\end{equation}
and the rest of commutators are zero. The Hamilton operator in terms
of the real fields has the form
\begin{equation} \label{40}
\begin{array}{l}
\displaystyle{%
   H=\int\! d{\bf x}\bigg[\frac{1}{2}\big(\pi_1^2+\pi_2^2\big)+    %
   \frac{1}{2}\Big(\!\big(\nabla\phi_1\big)^2+\big(\nabla\phi_2\big)^2\Big)+ %
   \frac{\kappa^2}{2}\big(\phi_1^2+\phi_2^2\big)+    %
   \frac{g_0}{4}\big(\phi_1^4+\phi_2^4+2\phi_1^2\phi_2^2\big)\bigg].    %
}
\end{array}
\end{equation}

The transition from the field operators to the particle operators in
the case of free linear fields is carried out in a known manner
[21,22]. For this purpose, one decomposes the field operator into
the frequency-positive and frequency-negative parts and passes to
the Fourier representation. The expansion coefficients have the
meaning of the creation and annihilation operators for
noninteracting particles. In the case of a nonlinear theory the
described recipe for introducing single-particle states cannot be
used, since now the Fourier transform coefficients do not have the
meaning of the operators of  creation and annihilation of states
with the necessary relativistic dispersion law.

It would be possible to single out in the initial Lagrangian the
part that is quadratic with respect to the field operators and
consider it, which is usually done, as the Lagrange function of free
particles. And then consider the rest of the Lagrangian as a
perturbation describing the interaction of particles. It should be
noted, however, that this seemingly ``obvious'' way of constructing
single-particle states is, in essence, an unspoken agreement. Such a
definition of the concept of ``particle'' is not necessary and the
only possible one. Moreover, it is incorrect in the case of states
with spontaneously broken symmetries, for the description of which
it is fundamentally important to take into account the effects
caused by the nonlinearity of the system. In particular, such
consideration does not make sense in the case when the mass
parameter $\kappa^2$ is negative, because then the free Lagrangian
describes objects with an ``irregular'' (tachyon) dispersion law.
The possibility of introducing tachyons into physics and giving them
a real status was widely discussed at one time. The need to
introduce such objects as tachyons has no serious justification. If
the field is described by nonlinear equations, then the presence in
the quadratic part of the Lagrangian of the ``wrong'' sign of the
squared mass does not at all indicate the existence of particles
with such exotic tachyon properties, although, as it was shown by
Goldstone [7], such a case describes a fully real physical situation. %

When constructing a nonlinear field theory, it should be remembered
that only the complete Lagrangian of the system has a physical
meaning. The decomposition of the complete Lagrangian into the sum
of the Lagrangians of ``noninteracting'' particles and their
interaction is obviously ambiguous. Indeed, in addition to some
decomposition $\Lambda=\Lambda_1+\Lambda_2$, one can start from
another decomposition $\Lambda=\Lambda_1'+\Lambda_2'$, where
$\Lambda_1'=\Lambda_1+\delta\Lambda$,
$\Lambda_2'=\Lambda_2-\delta\Lambda$, and $\delta\Lambda$ is some
operator addition. Fixing the way, in which it is constructed that
part of the Lagrangian which should describe single-particle states,
we thereby actually give a definition of the concept of ``free
particle'' in the framework of a nonlinear theory.

In the following, we will write the complete Lagrangian (28) as a sum %
\begin{equation} \label{41}
\begin{array}{l}
\displaystyle{%
   \Lambda=\Lambda_0+\Lambda_C,  %
}
\end{array}
\end{equation}
where $\Lambda_0$ is the Lagrangian containing all possible terms
not higher than quadratic in the field operators, including those
that violate the symmetry of the initial Lagrangian $\Lambda$. This
Lagrangian will be {\it by definition} considered as the Lagrangian
of ``noninteracting'' free particles. The second term $\Lambda_C$ is
the part of the Lagrangian containing all the other terms not
included in $\Lambda_0$, and it describes the interaction between
free particles. In the considered case of the complex scalar field,
the most general quadratic form of the Lagrangian has the form
\begin{equation} \label{42}
\begin{array}{l}
\displaystyle{%
   \Lambda_0=-\frac{\partial\phi^+}{\partial x_\mu}\frac{\partial\phi}{\partial x_\mu}  %
   -B\phi^{+2}-B^*\phi^2 - C\phi^+\phi -p\phi^+ -p^*\phi -V,  %
}
\end{array}
\end{equation}
where $B,C,p,V$ are $c$\,-number parameters, and $C$ and $V$ are
real. The interaction, or correlation, Lagrangian obviously has the form %
\begin{equation} \label{43}
\begin{array}{l}
\displaystyle{%
   \Lambda_C\equiv\Lambda-\Lambda_0=-g_0\phi^{+2}\phi^2-\big(\kappa^2-C\big)\phi^2  %
   +B\phi^{+2}+B^*\phi^2 +p\phi^+ +p^*\phi +V.  %
}
\end{array}
\end{equation}
As is seen, we have added and subtracted the operator addition %
$\delta\Lambda=-B\phi^{+2}-B^*\phi^2 -p\phi^+ -p^*\phi -V$ %
in the complete Lagrangian (28), without changing the Lagrangian
$\Lambda$ and, consequently, without making any approximation.
Parameters $B,C,p,V$ are assumed to be arbitrary for the time being,
and in what follows they should be chosen from some additional
considerations. Note that $\Lambda_0$ contains terms of the form
$\phi^{+2}, \phi^2, \phi^+, \phi $, which break the symmetry of the
original Lagrangian $\Lambda$ with respect to the phase
transformation of the field operators
\begin{equation} \label{44}
\begin{array}{l}
\displaystyle{%
   \phi \rightarrow \phi'=\phi\,e^{i\alpha},  %
}
\end{array}
\end{equation}
or in terms of the real and imaginary parts
\begin{equation} \label{45}
\begin{array}{l}
\displaystyle{%
   \phi_1' \rightarrow \phi_1\cos\alpha - \phi_2\sin\alpha,  \qquad %
   \phi_2' \rightarrow \phi_1\sin\alpha + \phi_2\cos\alpha,  %
}
\end{array}
\end{equation}
where $\alpha$ is the coordinate-independent real phase. A similar
statement is also true for $\Lambda_C$. Note that taking into
account pair correlations violating the phase symmetry, which is
ensured by the inclusion of terms with $\phi^{+2}, \phi^2$ into
Lagrangian (42), is important in nonrelativistic [3,4] and
relativistic [5,6] Fermi systems. For Bose systems with broken phase
symmetry [7], as a rule, pair correlations are not taken into
account, although, as was considered for nonrelativistic Bose
systems in [25,26] and shown for quantum relativistic fields in
[19,20] and in this work, they are equally important and lead to
qualitatively new results.

The energy-momentum tensor is also decomposed into the sum of two
terms, the main and correlation ones
\begin{equation} \label{46}
\begin{array}{l}
\displaystyle{%
   T_{\mu\nu}=T_{\mu\nu}^{(0)}+T_{\mu\nu}^{(C)},  %
}
\end{array}
\end{equation}
where
\begin{equation} \label{47}
\begin{array}{l}
\displaystyle{%
   T_{\nu\mu}^{(0)}=-\frac{\partial\phi^+}{\partial x_\mu}\frac{\partial\phi}{\partial x_\nu}  %
                    -\frac{\partial\phi^+}{\partial x_\nu}\frac{\partial\phi}{\partial x_\mu}  %
                    -\delta_{\nu\mu}\Lambda_0, %
}
\end{array}
\end{equation}
\vspace{-3mm}%
\begin{equation} \label{48}
\begin{array}{l}
\displaystyle{%
   T_{\nu\mu}^{(C)}= -\delta_{\nu\mu}\Lambda_C. %
}
\end{array}
\end{equation}
The full Hamiltonian is also split into the sum of two terms:
\begin{equation} \label{49}
\begin{array}{l}
\displaystyle{%
   H=H_0+H_C, %
}
\end{array}
\end{equation}
where in terms of the real fields
\begin{equation} \label{50}
\begin{array}{l}
\displaystyle{%
   H_0=\int\! d{\bf x}\bigg[\frac{1}{2}\big(\pi_1^2+\pi_2^2\big)+    %
   \frac{1}{2}\Big(\!\big(\nabla\phi_1\big)^2+\big(\nabla\phi_2\big)^2\Big)+ %
   \frac{B_1}{2}\phi_1^2 + \frac{B_2}{2}\phi_2^2 + C_0\phi_1\phi_2 +   %
   p_1\phi_1 + p_2\phi_2 + V \bigg],   %
}
\end{array}
\end{equation}
\vspace{-3mm}%
\begin{equation} \label{51}
\begin{array}{l}
\displaystyle{%
   H_C=\int\! d{\bf x}\bigg[\frac{g_0}{4}\big(\phi_1^4+\phi_2^4+2\phi_1^2\phi_2^2\big)+    %
   \frac{1}{2}\big(\kappa^2-B_1\big)\phi_1^2 + \frac{1}{2}\big(\kappa^2-B_2\big)\phi_2^2 -  %
   C_0\phi_1\phi_2 - p_1\phi_1 - p_2\phi_2 - V \bigg].
}%
\end{array}
\end{equation}
The coefficients entering into (50), (51) are related to the
coefficients in Lagrangians (42) and (43) by the relations
\begin{equation} \label{52}
\begin{array}{cc}
\displaystyle{%
   B_1=B+B^*+C, \qquad  B_2=-B-B^*+C, \qquad C_0=i\big(\!-B+B^*\big), %
}\vspace{3mm}\\ %
\displaystyle{\hspace{0mm}%
  p_1=\frac{(p+p^*)}{\sqrt{2}}, \qquad  p_2=-i\frac{(p-p^*)}{\sqrt{2}}. %
}%
\end{array}
\end{equation}
Since the full Hamiltonian is equally expressed in terms of both
Heisenberg and Schr\"{o}dinger operators, then $H_0$ and $H_C$ in
(50) and (51) can also be expressed in terms of the field operators
in these representations.

\section{Theory of ``free'' particles }\vspace{-0mm} %
Let us consider the main approximation, defined by the Hamiltonian
(50) written in terms of Schr\"{o}dinger operators. This
approximation describes a system of ``noninteracting'' or ``free'',
in the sense defined above, particles within the framework of the
self-consistent field model. Let us define the field operators in the interaction representation: %
\begin{equation} \label{53}
\begin{array}{l}
\displaystyle{%
   \hat{\phi}_i({\bf x},t)=e^{iH_0t}\phi_i({\bf x},0)e^{-iH_0t}, \quad      %
   \hat{\pi}_i({\bf x},t)=e^{iH_0t}\pi_i({\bf x},0)e^{-iH_0t} \quad  (i=1,2).      %
}%
\end{array}
\end{equation}
The simultaneous commutation relations for the operators (53) have
the same form as for the Heisenberg operators (39).  Obviously,
$H_0\equiv \hat{H}_0$ in the interaction representation is expressed
in terms of the operators (53) in the same way as in terms of
Schr\"{o}dinger operators:
\begin{equation} \label{54}
\begin{array}{l}
\displaystyle{%
   \hat{H}_0=\int\! d{\bf x}\bigg[\frac{1}{2}\big(\hat{\pi}_1^2+\hat{\pi}_2^2\big)+    %
   \frac{1}{2}\Big(\!\big(\nabla\hat{\phi}_1\big)^2+\big(\nabla\hat{\phi}_2\big)^2\Big)+ %
   \frac{B_1}{2}\hat{\phi}_1^2 + \frac{B_2}{2}\hat{\phi}_2^2 + C_0\hat{\phi}_1\hat{\phi}_2 +   %
   p_1\hat{\phi}_1 + p_2\hat{\phi}_2 + V \bigg].   %
}
\end{array}
\end{equation}
Later, when constructing the perturbation theory, we should also
write down the interaction Hamiltonian (51) in terms of the operators (53). %

Before reducing the Hamiltonian (54) to the diagonal form, it is
necessary to eliminate the linear terms in it by passing to the
``shifted'' operators $\hat{\psi}_i({\bf x},t)$:
\begin{equation} \label{55}
\begin{array}{l}
\displaystyle{%
   \hat{\phi}_i({\bf x},t)=\hat{\psi}_i({\bf x},t) + \chi_i, \qquad i=1,2.      %
}%
\end{array}
\end{equation}
Requiring the fulfillment of conditions
\begin{equation} \label{56}
\begin{array}{l}
\displaystyle{%
   B_1\chi_1+C_0\chi_2+p_1=0, \qquad B_2\chi_2+C_0\chi_1+p_2=0,  %
}%
\end{array}
\end{equation}
we arrive at the Hamiltonian, in which there are no terms linear in
the field operators:
\begin{equation} \label{57}
\begin{array}{l}
\displaystyle{%
   \hat{H}_0=\int\! d{\bf x}\bigg[\frac{1}{2}\big(\hat{\pi}_1^2+\hat{\pi}_2^2\big)+    %
   \frac{1}{2}\Big(\!\big(\nabla\hat{\psi}_1\big)^2+\big(\nabla\hat{\psi}_2\big)^2\Big)+ %
   \frac{B_1}{2}\hat{\psi}_1^2 + \frac{B_2}{2}\hat{\psi}_2^2 + C_0\hat{\psi}_1\hat{\psi}_2    %
   -\frac{B_1}{2}\chi_1^2 - \frac{B_2}{2}\chi_2^2 - C_0\chi_1\chi_2 + V \bigg].   %
}
\end{array}
\end{equation}
At the next stage, one should get rid of the cross terms in (57)
using the canonical Bogolyubov transformation
\begin{equation} \label{58}
\begin{array}{cc}
\displaystyle{%
   \hat{\psi}_1=\alpha\hat{u}_1+\beta\hat{u}_2, \qquad  \hat{\pi}_1=\alpha\hat{\upsilon}_1+\beta\hat{\upsilon}_2, %
}\vspace{3mm}\\ %
\displaystyle{\hspace{0mm}%
   \hat{\psi}_2=-\beta\hat{u}_1+\alpha\hat{u}_2, \qquad  \hat{\pi}_2=\beta\hat{\upsilon}_1-\alpha\hat{\upsilon}_2, %
}%
\end{array}
\end{equation}
where $\hat{\upsilon}_i\equiv\dot{\hat{u}}_i$, and the coefficients
in (58) are subject to the condition
\begin{equation} \label{59}
\begin{array}{cc}
\displaystyle{%
   \alpha^2+\beta^2 = 1, %
}%
\end{array}
\end{equation}
which for the operators of two introduced real fields ensures the
fulfillment of the commutation relations
\begin{equation} \label{60}
\begin{array}{l}
\displaystyle{%
   \big[\upsilon_1({\bf x},t),u_1({\bf x}',t)\big]=-i\delta({\bf x}-{\bf x}'), \qquad   %
   \big[\upsilon_2({\bf x},t),u_2({\bf x}',t)\big]=-i\delta({\bf x}-{\bf x}').   %
}
\end{array}
\end{equation}
By requiring that the cross terms drop out from the Hamiltonian
$\hat{H}_0$, we arrive at the relation
\begin{equation} \label{61}
\begin{array}{cc}
\displaystyle{%
   \alpha\beta\big(B_1-B_2\big)+\big(\alpha^2-\beta^2\big)C_0=0.  %
}%
\end{array}
\end{equation}
Without loss of generality, we will further assume $B_2-B_1>0$. From
(59) and (61) there follow the formulas:
\begin{equation} \label{62}
\begin{array}{cc}
\displaystyle{%
   \alpha^2=\frac{1}{2}\big(1+\sigma\big), \qquad  \beta^2=\frac{1}{2}\big(1-\sigma\big), \qquad \alpha\beta=\frac{C_0}{B_2-B_1}\,\sigma, \qquad %
   \sigma^2=\frac{\big(B_2-B_1\big)^2}{\big(B_2-B_1\big)^2+4C_0^2}.
}%
\end{array}
\end{equation}
As a result of the performed transformations, the Hamiltonian
$\hat{H}_0$ in terms of the operators $\hat{\upsilon}_i$ and
$\hat{u}_i$ takes the form
\begin{equation} \label{63}
\begin{array}{l}
\displaystyle{%
   \hat{H}_0=\int\! d{\bf x}\bigg[\frac{1}{2}\big(\hat{\upsilon}_1^2+\hat{\upsilon}_2^2\big)+    %
   \frac{1}{2}\Big(\!\big(\nabla\hat{u}_1\big)^2+\big(\nabla\hat{u}_2\big)^2\Big)+ %
   \frac{m_-^2}{2}\hat{u}_1^2 + \frac{m_+^2}{2}\hat{u}_2^2 %
   -\frac{B_1}{2}\chi_1^2-\frac{B_2}{2}\chi_2^2 - C_0\chi_1\chi_2 +V \bigg]. %
}%
\end{array}
\end{equation}
This Hamiltonian describes two free real fields of scalar particles
with nonzero masses
\begin{equation} \label{64}
\begin{array}{cc}
\displaystyle{%
   m_\pm^2=\frac{1}{2}\Big[B_1+B_2\pm\sqrt{\big(B_2-B_1\big)^2+4C_0^2} \,\,\Big]. %
}%
\end{array}
\end{equation}
The field operators in (63) obey the KGF equation of motion
\begin{equation} \label{65}
\begin{array}{cc}
\displaystyle{%
   \ddot{\hat{u}}_i - \Delta\hat{u}_i + m_i^2\hat{u}_i = 0 \qquad (i=1,2).  %
}%
\end{array}
\end{equation}
Here, and also below, we use the notation $m_1^2\equiv m_-^2, \,
m_2^2\equiv m_+^2$. The field operators can be expanded into the
Fourier integral
\begin{equation} \label{66}
\begin{array}{l}
\displaystyle{%
   \hat{u}_i(x)=\frac{1}{(2\pi)^{3\!/2}}\int\! \frac{d{\bf q}}{\sqrt{2\,q_{0i}}}    %
   \Big[a_i({\bf q})e^{iq_ix}+a_i^+({\bf q})e^{-iq_ix}\Big], %
}%
\end{array}
\end{equation}
\vspace{-3mm}%
\begin{equation} \label{67}
\begin{array}{l}
\displaystyle{%
   \hat{\upsilon}_i(x)=\frac{(-i)}{(2\pi)^{3\!/2}}\int\! d{\bf q}\,\sqrt{\frac{q_{0i}}{2} }\,   %
   \Big[a_i({\bf q})e^{iq_ix}-a_i^+({\bf q})e^{-iq_ix}\Big], %
}%
\end{array}
\end{equation}
where $q_i=\big({\bf q},iq_{0i}\big)$, $q_{0i}=\sqrt{{\bf
q}^2+m_i^2}$, $i=1,2$. The operators $a_i({\bf q}), \, a_i^+({\bf
q})$ have the meaning of operators of annihilation and creation of
particles with the mass $m_i$ and obey the well-known commutation relations %
\begin{equation} \label{68}
\begin{array}{l}
\displaystyle{%
   \big[a_i({\bf q}),a_{i'}^+({\bf q}')\big]=\delta({\bf q}-{\bf q}')\delta_{ii'}, \qquad   %
   \big[a_i({\bf q}),a_{i'}({\bf q}')\big]=0. %
}%
\end{array}
\end{equation}
Let us define the vacuum state vector $|0\rangle$ by the relation
$a_i({\bf q})|0\rangle=0$. This vector describes the state in which
there are no free particles defined by the method introduced above.
Then, obviously, we have the following vacuum averages
\begin{equation} \label{69}
\begin{array}{l}
\displaystyle{%
   \big\langle 0\big|\hat{u}_1\big|0\big\rangle =\big\langle 0\big|\hat{u}_2\big|0\big\rangle=0, \qquad   %
   \big\langle 0\big|\hat{u}_1\hat{u}_2\big|0\big\rangle=0, %
}%
\end{array}
\end{equation}
\vspace{-4mm}%
\begin{equation} \label{70}
\begin{array}{l}
\displaystyle{%
   \big\langle 0\big|\hat{\upsilon}_1\big|0\big\rangle =\big\langle 0\big|\hat{\upsilon}_2\big|0\big\rangle=0, \qquad   %
   \big\langle 0\big|\hat{\upsilon}_1\hat{\upsilon}_2\big|0\big\rangle=0, %
}%
\end{array}
\end{equation}
and also
\begin{equation} \label{71}
\begin{array}{l}
\displaystyle{%
   \big\langle 0\big|\hat{\psi}_1\big|0\big\rangle =\big\langle 0\big|\hat{\psi}_2\big|0\big\rangle=0.   %
}%
\end{array}
\end{equation}
It follows from the last formulas and relation (55) that
\begin{equation} \label{72}
\begin{array}{l}
\displaystyle{%
   \chi_1=\big\langle 0\big|\hat{\phi}_1\big|0\big\rangle, \qquad   %
   \chi_2=\big\langle 0\big|\hat{\phi}_2\big|0\big\rangle, \qquad   %
   \chi=\big\langle 0\big|\hat{\phi}\big|0\big\rangle,   %

}%
\end{array}
\end{equation}
and, consequently, the parameters $\chi_1$ and $\chi_2$ in (55) have
the meaning of the vacuum averages of the initial field operators.
The vacuum average of one field operator will be called the
\textit{one-particle} average.

An important role in the theory is played by taking into account the
vacuum average of the squares of the field operators (66), which can
be represented as
\begin{equation} \label{73}
\begin{array}{l}
\displaystyle{%
   \big\langle 0\big|\hat{u}_i^2(x)\big|0\big\rangle = %
   \frac{1}{2(2\pi)^3}\int\! \frac{d{\bf q}}{\sqrt{{\bf q}^2+m_i^2}} =   %
   -\lim_{\varepsilon\rightarrow +0}\frac{i}{(2\pi)^4}\int\! \frac{dq}{q^2+m_i^2-i\varepsilon}. %
}%
\end{array}
\end{equation}
The vacuum averages of products of two field operators will be
called \textit{pair} or \textit{two-particle} averages. The
well-known integral in formula (73) diverges at large momenta, so
that in order to obtain a finite result one has to introduce a
cutoff parameter $\lambda$. Using the relativistic-invariant cutoff, we find %
\begin{equation} \label{74}
\begin{array}{l}
\displaystyle{%
   \big\langle 0\big|\hat{u}_i^2(x)\big|0\big\rangle = %
   \frac{\lambda^2}{2\pi^2}\bigg[1-\frac{m_i^2}{\lambda^2}\ln\!\bigg(1+\frac{\lambda^2}{m_i^2}\bigg)\bigg]\equiv   %
   \frac{\lambda^2}{2\pi^2}\,f\bigg(\frac{m_i^2}{\lambda^2}\bigg). %
}%
\end{array}
\end{equation}
Here we introduce a function
\begin{equation} \label{75}
\begin{array}{l}
\displaystyle{%
   f(y)=1-y\ln\!\big(1+y^{-1}\big), %
}%
\end{array}
\end{equation}
that decreases monotonically from 1 at $y=0$\, to\, 0 at
$y\rightarrow\infty$.  Taking into account (58),\,(69), we also have
the following expressions for the pair vacuum averages
\begin{equation} \label{76}
\begin{array}{l}
\displaystyle{%
   \big\langle 0\big|\hat{\psi}_1^2\big|0\big\rangle = \alpha^2\big\langle 0\big|\hat{u}_1^2\big|0\big\rangle + \beta^2\big\langle 0\big|\hat{u}_2^2\big|0\big\rangle, %
}\vspace{3mm}\\ %
\displaystyle{\hspace{0mm}%
   \big\langle 0\big|\hat{\psi}_2^2\big|0\big\rangle = \beta^2\big\langle 0\big|\hat{u}_1^2\big|0\big\rangle + \alpha^2\big\langle 0\big|\hat{u}_2^2\big|0\big\rangle, %
}\vspace{3mm}\\ %
\displaystyle{\hspace{0mm}%
   \big\langle 0\big|\hat{\psi}_1\hat{\psi}_2\big|0\big\rangle = -\alpha\beta\big(\big\langle 0\big|\hat{u}_1^2\big|0\big\rangle -\big\langle 0\big|\hat{u}_2^2\big|0\big\rangle\big). %
}%
\end{array}
\end{equation}

So far, the parameters $B_1,B_2,C_0,p_1,p_2,V$ have not been fixed
in any way. We have to specify a method of finding these parameters.
By indicating this method, we will in fact complete the formulation
of the method of constructing single-particle states within the
framework of the developed self-consistent approach. We postulate
that the parameters $B_1,B_2,C_0,p_1,p_2,V$ must be found from the
requirement, consisting in that the approximating Hamiltonian
$\hat{H}_0$ be in some sense most close to the exact Hamiltonian
(40) or, in other words, the perturbation Hamiltonian be minimal. In
order to formulate this requirement quantitatively, we define the quantity %
\begin{equation} \label{77}
\begin{array}{l}
\displaystyle{\hspace{0mm}%
   J\equiv\big(\big\langle 0\big|H-H_0\big|0\big\rangle\big)^2\equiv\big(\big\langle 0\big|\hat{H}_C\big|0\big\rangle\big)^2. %
}%
\end{array}
\end{equation}
To calculate the average of the correlation Hamiltonian in (77), one
should use the relations
\begin{equation} \label{78}
\begin{array}{cc}
\displaystyle{%
   \big\langle 0\big|\hat{\psi}_1^4\big|0\big\rangle = 3\,\big(\big\langle 0\big|\hat{\psi}_1^2\big|0\big\rangle\big)^2, %
}\vspace{3mm}\\ %
\displaystyle{%
   \big\langle 0\big|\hat{\psi}_1^2\hat{\psi}_2^2\big|0\big\rangle = 2\,\big(\big\langle 0\big|\hat{\psi}_1\hat{\psi}_2\big|0\big\rangle\big)^2 + %
   \big\langle 0\big|\hat{\psi}_1^2\big|0\big\rangle\big\langle 0\big|\hat{\psi}_2^2\big|0\big\rangle. %
}%
\end{array}
\end{equation}
Let us introduce the following notation for the vacuum averages of
the products of two operators:
\begin{equation} \label{79}
\begin{array}{cc}
\displaystyle{%
   \rho_1=\big\langle 0\big|\hat{\phi}_1^2\big|0\big\rangle, \qquad %
   \rho_2=\big\langle 0\big|\hat{\phi}_2^2\big|0\big\rangle, \qquad %
   \rho_{12}=\rho_{21}=\big\langle 0\big|\hat{\phi}_1\hat{\phi}_2\big|0\big\rangle,  %
}%
\end{array}
\end{equation}
\vspace{-4mm}%
\begin{equation} \label{80}
\begin{array}{cc}
\displaystyle{%
   \tilde{\rho}_1=\big\langle 0\big|\hat{\psi}_1^2\big|0\big\rangle, \qquad %
   \tilde{\rho}_2=\big\langle 0\big|\hat{\psi}_2^2\big|0\big\rangle, \qquad %
   \tilde{\rho}_{12}=\tilde{\rho}_{21}=\big\langle 0\big|\hat{\psi}_1\hat{\psi}_2\big|0\big\rangle.  %
}%
\end{array}
\end{equation}
There is an obvious connection between these pair vacuum averages
\begin{equation} \label{81}
\begin{array}{cc}
\displaystyle{%
   \rho_1=\tilde{\rho}_1+\chi_1^2, \qquad %
   \rho_2=\tilde{\rho}_2+\chi_2^2, \qquad %
   \rho_{12}=\tilde{\rho}_{12}+\chi_1\chi_2.   %
}%
\end{array}
\end{equation}
Taking into account relations (78)\,--\,(81), we find the vacuum
average of the density of the correlation Hamiltonian
\begin{equation} \label{82}
\begin{array}{ll}
\displaystyle{\hspace{0mm}%
   \frac{\big\langle 0\big|\hat{H}_C\big|0\big\rangle}{\Omega}= %
   \frac{g_0}{4}\Big[3\rho_1^2+3\rho_2^2+4\rho_{12}^2+2\rho_1\rho_2-2\big(\chi_1^2+\chi_2^2\big)^2\Big]- %
}\vspace{2mm}\\ %
\displaystyle{\hspace{17mm}%
   -\frac{1}{2}\big(B_1-\kappa^2\big)\rho_1 -\frac{1}{2}\big(B_2-\kappa^2\big)\rho_2 %
   -C_0\rho_{12} - p_1\chi_1 - p_2\chi_2 - V,  %
}%
\end{array}
\end{equation}
where $\Omega$ is the spatial volume. Then from the conditions
\begin{equation} \label{83}
\begin{array}{cc}
\displaystyle{%
   \frac{\partial J}{\partial\rho_1}=\frac{\partial J}{\partial\rho_2}=  %
   \frac{\partial J}{\partial\rho_{12}}=\frac{\partial J}{\partial\chi_1}=\frac{\partial J}{\partial\chi_2}=0 %
}%
\end{array}
\end{equation}
we find expressions for the coefficients in the free Hamiltonian
$\hat{H}_0$, which are determined in terms of single-particle (72)
and pair (79) vacuum averages
\begin{equation} \label{84}
\begin{array}{ccc}
\displaystyle{\hspace{0mm}%
  B_1=\kappa^2+g_0\big(3\rho_1+\rho_2\big), \qquad B_2=\kappa^2+g_0\big(3\rho_2+\rho_1\big),  %
}\vspace{3mm}\\ %
\displaystyle{\hspace{17mm}%
  C_0=2g_0\rho_{12},  %
}\vspace{3mm}\\ %
\displaystyle{\hspace{17mm}%
  p_1=-2g_0\chi_1\big(\chi_1^2+\chi_2^2\big),  \qquad  p_2=-2g_0\chi_2\big(\chi_1^2+\chi_2^2\big).    
}%
\end{array}
\end{equation}
This implies $B_2-B_1=2g_0(\rho_2-\rho_1)$. Since we assumed that
$B_2-B_1>0$, then it should also be $\rho_2-\rho_1 > 0$.

The quantity $J$ (77), which characterizes the deviation of the
approximating Hamiltonian from the exact one, takes a minimum value
under the fulfillment of the condition
\begin{equation} \label{85}
\begin{array}{c}
\displaystyle{\hspace{0mm}%
   \big\langle 0\big|\hat{H}_C\big|0\big\rangle=0, %
}%
\end{array}
\end{equation}
which allows to determine the parameter $V$:
\begin{equation} \label{86}
\begin{array}{cc}
\displaystyle{\hspace{0mm}%
   V= \frac{g_0}{4}\Big[3\rho_1^2+3\rho_2^2+4\rho_{12}^2+2\rho_1\rho_2-2\big(\chi_1^2+\chi_2^2\big)^2\Big]- %
}\vspace{2mm}\\ %
\displaystyle{\hspace{5mm}%
   -\frac{1}{2}\big(B_1-\kappa^2\big)\rho_1 -\frac{1}{2}\big(B_2-\kappa^2\big)\rho_2 %
   -C_0\rho_{12} - p_1\chi_1 - p_2\chi_2 .  %
}%
\end{array}
\end{equation}
Note that the condition (85) is completely natural and ensures
turning to zero of the interaction energy in the absence of
particles. As we can see, taking into account the non-operator term
$V$ in the approximating Lagrangian (42) is fundamentally important,
since it allows to correctly determine the shift in the energy of
the ground state at transition to a self-consistent description.

Substituting (84) into (56), we find the nonlinear equations that
determine the one-particle vacuum averages (72):
\begin{equation} \label{87}
\begin{array}{cc}
\displaystyle{\hspace{0mm}%
   \big[\kappa^2+g_0\big(3\rho_1+\rho_2\big)\big]\chi_1 + 2g_0\rho_{12}\chi_2-2g_0\chi_1\big(\chi_1^2+\chi_2^2\big)=0,  %
}%
\end{array}
\end{equation}
\vspace{-4mm}
\begin{equation} \label{88}
\begin{array}{cc}
\displaystyle{\hspace{0mm}%
   \big[\kappa^2+g_0\big(3\rho_2+\rho_1\big)\big]\chi_2 + 2g_0\rho_{12}\chi_1-2g_0\chi_2\big(\chi_1^2+\chi_2^2\big)=0,  %
}%
\end{array}
\end{equation}
where, as before, $g_0\equiv g\big/4!$. Taking into account formulas
for the coefficients (84), for the parameters of the canonical
transformation (62) we find the relations
\begin{equation} \label{89}
\begin{array}{ll}
\displaystyle{\hspace{0mm}%
  \sigma=\frac{\rho_1-\rho_2}{Z}, \qquad \alpha\beta = -\frac{\rho_{12}}{Z}, %
}%
\end{array}
\end{equation}
where it is introduced a notation for the positive quantity
\begin{equation} \label{90}
\begin{array}{ll}
\displaystyle{\hspace{0mm}%
  Z\equiv\sqrt{(\rho_1-\rho_2)^2+4\rho_{12}^2}\,\,.  %
}%
\end{array}
\end{equation}
The substitution of the coefficients from (84) into (64) makes it
possible to express the masses of scalar particles in terms of the
vacuum averages
\begin{equation} \label{91}
\begin{array}{ll}
\displaystyle{\hspace{0mm}%
  m_\pm^2\equiv m_{2,1}^2=\kappa^2+g_0\big[2(\rho_1+\rho_2)\pm Z\big].   %
}%
\end{array}
\end{equation}
Taking also into account (74) and (76), we obtain the following
formulas for the quadratic vacuum averages of the overcondensate
field operators (80):
\begin{equation} \label{92}
\begin{array}{cc}
\displaystyle{\hspace{0mm}%
   \tilde{\rho}_1=\frac{\lambda^2}{4\pi^2}\bigg[\big(\tilde{f}_1+\tilde{f}_2\big)+\frac{(\rho_1-\rho_2)}{Z}\big(\tilde{f}_1-\tilde{f}_2\big)\bigg], %
}\vspace{2mm}\\ %
\displaystyle{\hspace{0mm}%
   \tilde{\rho}_2=\frac{\lambda^2}{4\pi^2}\bigg[\big(\tilde{f}_1+\tilde{f}_2\big)-\frac{(\rho_1-\rho_2)}{Z}\big(\tilde{f}_1-\tilde{f}_2\big)\bigg], %
}\vspace{2mm}\\ %
\displaystyle{\hspace{0mm}%
   \tilde{\rho}_{12}=\frac{\lambda^2}{2\pi^2}\frac{\rho_{12}}{Z}\big(\tilde{f}_1-\tilde{f}_2\big), %
}%
\end{array}
\end{equation}
where $\tilde{f}_1\equiv f\big(\tilde{m}_1^2\big)$,
$\tilde{f}_2\equiv f\big(\tilde{m}_2^2\big)$, and
$\tilde{m}_i^2=m_i^2\big/\lambda^2$. Formulas (81),\,(87),\,(88) and
(90)\,--\,(92) form a system of nonlinear equations for finding the
vacuum averages and particle masses. These equations can have
solutions of various symmetries, which will be discussed in the next section. %

\section{Masses of particles in different phases}\vspace{-0mm} %
The field state is characterized by the normal phase-invariant pair
vacuum average
\begin{equation} \label{93}
\begin{array}{ll}
\displaystyle{\hspace{0mm}%
  \big\langle 0\big|\hat{\phi}^+\hat{\phi}\big|0\big\rangle=\frac{1}{2}\big(\rho_1+\rho_2\big),   %
}%
\end{array}
\end{equation}
as well as by the anomalous single-particle
\begin{equation} \label{94}
\begin{array}{cc}
\displaystyle{\hspace{0mm}%
  \big\langle 0\big|\hat{\phi}\big|0\big\rangle=\frac{1}{\sqrt{2}}\big(\chi_1+i\chi_2\big), \qquad  %
  \big\langle 0\big|\hat{\phi}^+\big|0\big\rangle=\frac{1}{\sqrt{2}}\big(\chi_1-i\chi_2\big)  %
}%
\end{array}
\end{equation}
and anomalous pair vacuum averages
\begin{equation} \label{95}
\begin{array}{cc}
\displaystyle{\hspace{0mm}%
  \big\langle 0\big|\hat{\phi}^2\big|0\big\rangle=\frac{1}{2}\big(\rho_1-\rho_2+2i\rho_{12}\big), \qquad  %
  \big\langle 0\big|\hat{\phi}^{+2}\big|0\big\rangle=\frac{1}{2}\big(\rho_1-\rho_2-2i\rho_{12}\big),   %
}%
\end{array}
\end{equation}
violating the symmetry of the state with respect to the phase
transformations (44),\,(45), which, according to the terminology
adopted in statistical physics, are order parameters.  The averages
of the product of a larger number of operators are expressed in
terms of the averages (93)\,--\,(95).

\newpage %
\hspace{60mm}%
5.1. \textit{Symmetric state} ($s$\,-\textit{phase}) %

Equations (81),\,(87),\,(88) and (90)\,--\,(92) admit the solution
$\rho_1=\rho_2\equiv\rho=\frac{\lambda^2}{2\pi^2}f\big(\tilde{m}^2\big)$
and $\chi_1=\chi_2=\rho_{12}=0$, for which all anomalous averages
(94),\,(95) violating the phase symmetry of the state vanish. We
will call such a phase-invariant state of the field the symmetric
phase ($s$\,-phase). In this case the complex field is equivalent to
two real fields, the particles of which according to (90),\,(91)
have the same mass $\tilde{m}_s^2$:
\begin{equation} \label{96}
\begin{array}{ll}
\displaystyle{\hspace{0mm}%
  \tilde{m}_s^2=\tilde{\kappa}^2 +2g_*f\big(\tilde{m}_s^2\big),
}%
\end{array}
\end{equation}
where $g_*\equiv g_0\big/\pi^2$,
$\tilde{\kappa}^2\equiv\kappa^2\big/\lambda^2$. The dependence of
the mass $\tilde{m}_s^2$ on the mass parameter $\tilde{\kappa}^2$ at
the fixed $g_*$  is shown in Fig.\,1 (line $aa_1$). For large
positive values of $\tilde{\kappa}^2$ the dependence of the particle
mass on this parameter is close to linear: $\tilde{m}_s^2\approx\tilde{\kappa}^2$. %
As the parameter $\tilde{\kappa}^2$ decreases, the mass $\tilde{m}_s^2$ %
decreases monotonically and vanishes at $\tilde{\kappa}^2=-2g_*$.
For $\tilde{\kappa}^2<-2g_*$, $s$\,-phase cannot exist.
\vspace{0mm} %
\begin{figure}[h!]
\vspace{-0mm}  \hspace{0mm}
\includegraphics[width = 6.7cm]{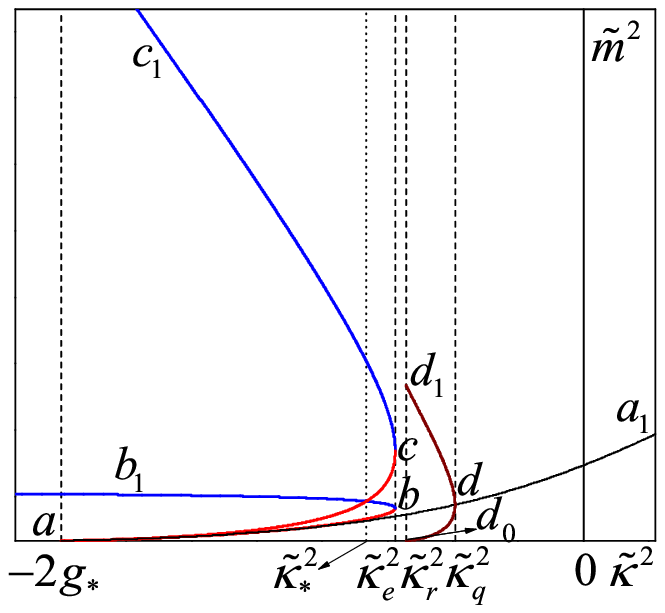} 
\vspace{-4mm} %
\caption{\label{fig01} 
Dependences of the particle masses on the mass parameter
$\tilde{m}^2=\tilde{m}^2\big(\tilde{\kappa}^2\big)$ at $g_*=4.0$. \newline %
($aa_1$): $\tilde{m}_s^2$ -- $s$\,-phase; %
($bb_1$): $\tilde{m}_{h1}^2$, ($cc_1$): $\tilde{m}_{h2}^2$ -- $b_h$\,-phase; %
($ab$): $\tilde{m}_{l1}^2$, ($ac$): $\tilde{m}_{l2}^2$ -- $b_l$\,-phase; %
($d_0d$): $\tilde{m}_{p1}^2$, ($dd_1$): $\tilde{m}_{p2}^2$ -- $p$\,-phase. \newline %
$\tilde{\kappa}_*^2=-3.33,\, \tilde{\kappa}_e^2=-2.88,\, \tilde{\kappa}_r^2=-2.72,\, \tilde{\kappa}_q^2=-1.97$. %
}%
\end{figure}

\hspace{60mm}%
5.2. \textit{Asymmetric state} I ($b$\,-\textit{phases}) %

Let us consider states with broken phase symmetry for the case when
both the single-particle (94) and pair anomalous vacuum averages
(95) are nonzero. We call such states $b$\,-phase. Equations
(87),\,(88) allow to express the single-particle averages in terms
of the pair averages:
\begin{equation} \label{97}
\begin{array}{cc}
\displaystyle{\hspace{0mm}%
  \chi_1^2=\frac{(d+Z)}{4}\bigg[1+\frac{(\rho_1-\rho_2)}{Z}\bigg], \qquad %
  \chi_2^2=\frac{(d+Z)}{4}\bigg[1-\frac{(\rho_1-\rho_2)}{Z}\bigg], \qquad  %
  \chi_1\chi_2=\frac{(d+Z)}{2Z}\rho_{12}, %
}%
\end{array}
\end{equation}
where $\displaystyle{d\equiv\frac{\kappa^2}{g_0}}+2(\rho_1+\rho_2)$. %
Taking into account these relations, from equations (92) we find
\begin{equation} \label{98}
\begin{array}{cc}
\displaystyle{\hspace{0mm}%
  Z=-\frac{\kappa^2}{g_0}-\frac{\lambda^2}{\pi^2}\big(\tilde{f}_1+\tilde{f}_2\big), \qquad %
  \rho_1+\rho_2=-\frac{\kappa^2}{g_0}-\frac{\lambda^2}{\pi^2}\tilde{f}_2. %
}%
\end{array}
\end{equation}
Substituting expressions (98) into (91), we arrive at a system of
equations for the particle masses in the considered phase
\begin{equation} \label{99}
\begin{array}{cc}
\displaystyle{\hspace{0mm}%
  \tilde{m}_1^2=g_*\big[f\big(\tilde{m}_1^2\big)-f\big(\tilde{m}_2^2\big)\big], %
}\vspace{3mm}\\ %
\displaystyle{\hspace{0mm}%
  \tilde{m}_2^2=-2\tilde{\kappa}^2-g_*\big[f\big(\tilde{m}_1^2\big)+3f\big(\tilde{m}_2^2\big)\big],
}%
\end{array}
\end{equation}
where, as above, $\tilde{\kappa}^2\equiv\kappa^2\big/\lambda^2, \, g_*\equiv g_0\big/\pi^2$. %

To solve these equations, it is convenient to introduce the quantities %
$t_1\equiv\ln\!\big(1+\tilde{m}_1^{-2}\big)$, $t_2\equiv\ln\!\big(1+\tilde{m}_2^{-2}\big)$, %
in terms of which the masses are expressed as
\begin{equation} \label{100}
\begin{array}{ll}
\displaystyle{\hspace{0mm}%
  \tilde{m}_1^2=\frac{1}{e^{t_1}-1}, \qquad \tilde{m}_2^2=\frac{1}{e^{t_2}-1}, %
}%
\end{array}
\end{equation}
as well as the function of these quantities
\begin{equation} \label{101}
\begin{array}{ll}
\displaystyle{\hspace{0mm}%
  u(t_1,t_2)\equiv\frac{(2\upsilon+t_1)}{e^{t_1}-1} + \frac{3t_2}{e^{t_2}-1}%
  -2\upsilon\bigg(\frac{1}{e^{t_1}-1}+\frac{1}{e^{t_2}-1}\bigg)-4, %
}%
\end{array}
\end{equation}
where $\upsilon\equiv 1\big/2g_*$. In this notation equations (99) take the form %
\begin{equation} \label{102}
\begin{array}{ll}
\displaystyle{\hspace{0mm}%
  \frac{t_1+2\upsilon}{e^{t_1}-1}=\frac{t_2}{e^{t_2}-1}\equiv p, %
}%
\end{array}
\end{equation}
\vspace{-3mm}%
\begin{equation} \label{103}
\begin{array}{ll}
\displaystyle{\hspace{0mm}%
  4\upsilon\tilde{\kappa}^2=u(t_1,t_2).  %
}%
\end{array}
\end{equation}
Since equation (102) must be satisfied for all permissible values of
$t_1,\,t_2$, we have equated its left and right parts to one
parameter $p$, and consider the quantities $t_1=t_1(p),\,t_2=t_2(p)$
as functions of $p$. Then the right side of equation (103)
$u\big(t_1(p),t_2(p)\big)\equiv \tilde{u}(p)$ is also a function of
$p$. Since $0<t_1<\infty,\,0<t_2<\infty$, then the parameter $p$
changes in the range $0<p<1$.

Let us analyze the properties of the function $\tilde{u}(p)$, the
form of which is shown in Fig.\,2. In the limit $p\rightarrow 0$  we
have $\tilde{u}(0)=-4$, and the derivative $\tilde{u}'(0)=4$ is
positive. In the other limit $p\rightarrow 1$ the function
$\tilde{u}\rightarrow -\infty$, and its derivative is negative and
$\tilde{u}'\rightarrow -\infty$. At some value $p_e$, the function
$\tilde{u}(p)$ reaches a maximum value $\tilde{u}_e$. These
quantities are found from the equations:
\begin{equation} \label{104}
\begin{array}{cc}
\displaystyle{\hspace{0mm}%
  2-\upsilon\bigg(\frac{1}{e^{-t_{e1}}+2\upsilon+t_{e1}-1}+\frac{1}{e^{-t_{e2}}+t_{e2}-1}\bigg)=0, \qquad %
  p_e=\frac{2\upsilon+t_{e1}}{e^{t_{e1}}-1}=\frac{t_{e2}}{e^{t_{e2}}-1}, %
}\vspace{3mm}\\ %
\displaystyle{\hspace{0mm}%
  \tilde{u}_e=4p_e\bigg[1-\frac{\upsilon}{2}\bigg(\frac{1}{2\upsilon+t_{e1}}+\frac{1}{t_{e2}}\bigg)\bigg]-4.    %
}%
\end{array}
\end{equation}
For $p>p_e$ the function $\tilde{u}(p)$ decreases, and at the point
$p_0$, defined by the formulas
\begin{equation} \label{105}
\begin{array}{cc}
\displaystyle{\hspace{0mm}%
   2-\upsilon\bigg(\frac{1}{2\upsilon+t_{10}}+\frac{1}{t_{20}}\bigg)=0, \qquad %
   p_0=\frac{2\upsilon+t_{10}}{e^{t_{10}}-1}=\frac{t_{20}}{e^{t_{20}}-1}, %
}%
\end{array}
\end{equation}
becomes equal to the value of the function at the origin $u(p_0)=u(0)=-4$. %
\vspace{0mm} %
\begin{figure}[h!]
\vspace{-2mm}  \hspace{0mm}
\includegraphics[width = 7.7cm]{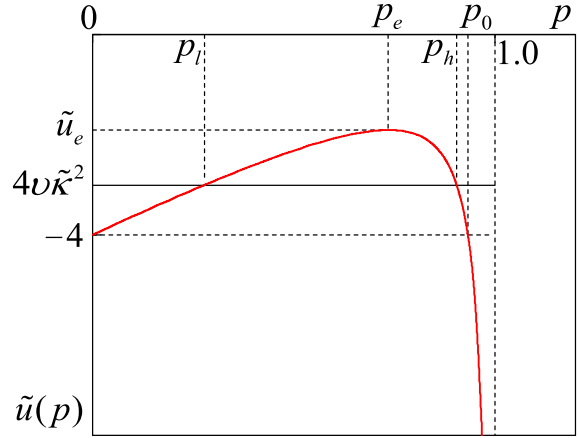} 
\vspace{-0mm} %
\caption{\label{fig02} 
The function $\tilde{u}(p)$ at $g_*=2$. Numerical values at characteristic points: %
$(p_e,\tilde{u}_e)=(0.74,-1.90)$,\,$(p_0,\tilde{u}_0)=(0.93,-4)$. %
}%
\end{figure}

The dependence of the particle masses in both $b$\,-phases on the
parameter $\tilde{\kappa}^2$ at fixed $g_*$ is shown in Fig.\,1. For
$\tilde{\kappa}^2>\tilde{\kappa}_e^2\equiv g_*\big/2\tilde{u}_e$
equations (102),\,(103) have no solutions. The solution with
$\tilde{m}_{e2}^2>\tilde{m}_{e1}^2$ appears at
$\tilde{\kappa}^2=\tilde{\kappa}_e^2$ (points $c$ and $b$ in
Fig.\,1). In the interval
$-2g_*\leq\tilde{\kappa}^2<\tilde{\kappa}_e^2$ with parameters
$p_l<p_h$, there are two solutions with
$\tilde{m}_{l2}^2>\tilde{m}_{l1}^2$ (lines $ac$,\,$ab$) and with
$\tilde{m}_{h2}^2>\tilde{m}_{h1}^2$ (lines $cc_1$,\,$bb_1$). The
phases with broken symmetry corresponding to these solutions are
denoted as $b_l$ and $b_h$. At the point $\tilde{\kappa}^2=-2g_*$
itself, we have $p_l=0,\, p_h=p_0$ and the masses
$\tilde{m}_{l2}^2=\tilde{m}_{l1}^2=0$ vanish, while the other two
masses take on the finite values $\tilde{m}_{h10}^2$ and
$\tilde{m}_{h20}^2$. For $\tilde{\kappa}^2<-2g_*$, only one solution
remains with $\tilde{m}_{h1}^2$ and $\tilde{m}_{h2}^2$ (lines
$cc_1$,\,$bb_1$). In the limit $\tilde{\kappa}^2\rightarrow -\infty$
these masses tend to limit values:
\begin{equation} \label{106}
\begin{array}{cc}
\displaystyle{\hspace{0mm}%
   \tilde{m}_{h1\infty}^2=\frac{1}{e^{t_{h1\infty}}-1}, \qquad  %
   \tilde{m}_{h2\infty}^2=-2\tilde{\kappa}^2,           \qquad  %
   \frac{2\upsilon+t_{h1\infty}}{e^{t_{h1\infty}}-1}=1.  %
}%
\end{array}
\end{equation}

\newpage
\hspace{60mm}%
5.3. \textit{Asymmetric state} II ($p$\,-\textit{phase}) %

In addition to considered in 5.2 phases, where the phase symmetry is
broken due to the existence of the non-zero anomalous averages (94)
and (95), a state proves to be possible in which $\chi_1=\chi_2=0$,
and the phase symmetry is broken only due to the non-zero quadratic
anomalous averages $\big\langle
0\big|\hat{\phi}^2\big|0\big\rangle,\, \big\langle
0\big|\hat{\phi}^{+2}\big|0\big\rangle$. Consider this state, which
we call $p$\,-phase. In this state $\rho_1=\tilde{\rho}_1,\,
\rho_2=\tilde{\rho}_2,\, \rho_{12}=\tilde{\rho}_{12}$. From formulas
(92) we find:
\begin{equation} \label{107}
\begin{array}{cc}
\displaystyle{\hspace{0mm}%
  \rho_1+\rho_2=\frac{\lambda^2}{2\pi^2}\big(\tilde{f}_1+\tilde{f}_2\big), \qquad %
  Z=\frac{\lambda^2}{2\pi^2}\big(\tilde{f}_1-\tilde{f}_2\big). %
}%
\end{array}
\end{equation}
Substituting (107) into (91), we find equations for the masses
\begin{equation} \label{108}
\begin{array}{cc}
\displaystyle{\hspace{0mm}%
  \tilde{m}_{p1}^2=\tilde{\kappa}^2+\frac{g_*}{2}\big[f\big(\tilde{m}_{p1}^2\big)+3f\big(\tilde{m}_{p2}^2\big)\big], %
}\vspace{3mm}\\ %
\displaystyle{\hspace{0mm}%
  \tilde{m}_{p2}^2=\tilde{\kappa}^2+\frac{g_*}{2}\big[3f\big(\tilde{m}_{p1}^2\big)+f\big(\tilde{m}_{p2}^2\big)\big]. %
}%
\end{array}
\end{equation}
The system of equations (108) at fixed $g_*$ has a solution in the
region $\tilde{\kappa}_r^2\le\tilde{\kappa}^2\le\tilde{\kappa}_q^2$.
The region of existence of this phase is narrow at small $g_*$ and
rapidly expands with an increase in this parameter. At
$\tilde{\kappa}^2=\tilde{\kappa}_q^2$ both masses are the same and
equal to $\tilde{m}_{q}^2$. The quantities $\tilde{\kappa}_q^2$ and
$\tilde{m}_{q}^2$ are found from the system of equations
\begin{equation} \label{109}
\begin{array}{cc}
\displaystyle{\hspace{0mm}%
  \frac{1}{g_*}-\ln\!\big(1+\tilde{m}_{q}^{-2}\big)+\frac{1}{1+\tilde{m}_{q}^2}=0, \qquad %
  \tilde{\kappa}_q^2=3\tilde{m}_{q}^2-\frac{2g_*}{1+\tilde{m}_{q}^2}. %
}%
\end{array}
\end{equation}
At $\tilde{\kappa}^2=\tilde{\kappa}_r^2=-2g_*+3\Big/2\big(e^{1\!/g_*}-1\big)$ %
the smaller mass $\tilde{m}_{p1}^2$ (line $d_0d$) vanishes, and the
larger mass $\tilde{m}_{p2}^2$ (line $dd_1$) takes the finite value
$\tilde{m}_{p2}^2\equiv\tilde{m}_r^2=\big(e^{1\!/g_*}-1\big)^{-1}$ %
(Fig.\,1). Characteristically, for the squared dimensionless masses
normalized to the cutoff parameter $\tilde{m}^2\equiv m^2\big/\lambda^2$ %
and $\tilde{\kappa}^2\equiv \kappa^2\big/\lambda^2$, equations
(96),\,(99), and (108) do not contain explicitly the cutoff
parameter itself.

It should be noted that there is no solution to equations
(87),\,(88),\,(92), in which the phase symmetry would be violated
only due to the single-particle average $\chi$ and at that the
anomalous pair vacuum averages (95) would be equal to zero. Only if
we formally neglect the quadratic vacuum averages of the
overcondensate field operators, setting
$\tilde{\rho}_1=\tilde{\rho}_2=\tilde{\rho}_{12}=0$, then we have
$\rho_1=\chi_1^2,\,\rho_2=\chi_2^2,\,\rho_{12}=\chi_1\chi_2$. %
Substituting these expressions into (91), we obtain
\begin{equation} \label{110}
\begin{array}{cc}
\displaystyle{\hspace{0mm}%
  m_+^2\equiv m_2^2=\kappa^2+3g_0\big(\chi_1^2+\chi_2^2\big), \qquad %
  m_-^2\equiv m_1^2=\kappa^2+g_0\big(\chi_1^2+\chi_2^2\big).  %

}%
\end{array}
\end{equation}
Equations (87),\,(88) in this case imply the relation
\begin{equation} \label{111}
\begin{array}{cc}
\displaystyle{\hspace{0mm}%
  \chi_1^2+\chi_2^2=-\frac{\kappa^2}{g_0}, %
}%
\end{array}
\end{equation}
so that, neglecting the quadratic vacuum averages (80), we arrive at
the Goldstone result [7]
\begin{equation} \label{112}
\begin{array}{cc}
\displaystyle{\hspace{0mm}%
  m_+^2 = -2\kappa^2, \qquad %
  m_-^2 = 0.   %
}%
\end{array}
\end{equation}

Thus, we showed in this section that, depending on the values of the
parameters $\kappa^2$ and $g$ entering into the initial Lagrangian
(28), the quantum complex scalar field can exist either in the
symmetric state, where the phase symmetry is not broken
($s$\,-phase), or in the states with broken phase symmetry
($b_l$,\,$b_h$\,-\,phases or $p$\,-phase). In $s$\,-phase all
particles have the same mass, and in $b$\,-phases and $p$\,-phase
there are particles of two types -- ``heavy'' and ``light'' ones,
having a finite mass. If the existence of several phases is possible
for the same values of the parameters $\kappa^2$ and $g$, then the
phase with the lowest energy of the vacuum state is realized.

\section{Vacuum state}\vspace{-0mm} %
The important role of the vacuum state for understanding the laws of
nature was recognized quite a long time ago (see, for example,
[27]). In Section 4 the vacuum state vector, in which there are no
particles, was defined by the condition $a_i({\bf q})|0\rangle=0$,
where $a_i({\bf q})$ is the operator coefficient in the Fourier
expansions (66),\,(67) which have the meaning of the annihilation
operator for a free scalar particle with the mass $m_i$. It is
natural to call this vacuum the vacuum of particles. The energy
operator (63), written in terms of the particle creation and
annihilation operators, has the form
\begin{equation} \label{113}
\begin{array}{cc}
\displaystyle{%
   \hat{H}_0=\sum_{i=1,2}\int\! d{\bf q}\, q_{0i}\,a_i^+({\bf q})a_i({\bf q}) +    %
   \frac{1}{2}\int\! d{\bf q}d{\bf q}'\,q_{0i}\delta({\bf q}-{\bf q}')\delta({\bf q}-{\bf q}')\, +  %
}\vspace{3mm}\\ %
\displaystyle{\hspace{0mm}%
  +\,\Omega\,\frac{g_0}{2}\bigg[\big(\chi_1^2+\chi_2^2\big)^2 - \frac{1}{2}\big(3\rho_1^2+3\rho_2^2+4\rho_{12}^2+2\rho_{1}\rho_{2}\big)\bigg].   %
}%
\end{array}
\end{equation}
The second term in (113) can be calculated if we take account
that $\delta(0)=\Omega\big/\!(2\pi)^3$, $\Omega$ is the volume. Then we find %
\begin{equation} \label{114}
\begin{array}{cc}
\displaystyle{%
  A_i\equiv\frac{1}{2}\int\! d{\bf q}d{\bf q}'\, q_{0i}\delta({\bf q}-{\bf q}')\delta({\bf q}-{\bf q}') =  %
  \frac{\Omega}{2(2\pi)^3}\int\! d{\bf q}\sqrt{{\bf q}^2+m_i^2}\equiv\frac{\Omega J_i}{2(2\pi)^3}, %
}%
\end{array}
\end{equation}
where $J_i\equiv\int\! d{\bf q}\sqrt{{\bf q}^2+m_i^2}$, $i=1,2$.
Above we used the cutoff of the divergent integral at large momenta
(74). The divergent integral $J_i$ can be calculated by regularizing
it in the same way as was done in calculating the vacuum average
integral in (73). Differentiating $J_i$ by $m_i^2$, we have
\begin{equation} \label{115}
\begin{array}{cc}
\displaystyle{%
  \frac{dJ_i}{dm_i^2}=\frac{1}{2}\int\!d{\bf q}\,\big({\bf q}^2+m_i^2\big)^{-1\!/2} =  %
  4\pi\lambda^2\big[1-\tilde{m}_i^2\ln\!\big(1+\tilde{m}_i^{-2}\big)\big] \equiv %
  4\pi\lambda^2 f\big(\tilde{m}_i^2\big). %
}%
\end{array}
\end{equation}
Integrating the last relation, we obtain
\begin{equation} \label{116}
\begin{array}{cc}
\displaystyle{%
  A_i=\Omega\,\varepsilon_0\big[-\tilde{m}_i^4\ln\!\big(1+\tilde{m}_i^{-2}\big) +  %
  \ln\!\big(1+\tilde{m}_i^{2}\big) + \tilde{m}_i^{2} + C_i \big], %
}%
\end{array}
\end{equation}
where $C_i$ is the constant of integration, which does not depend on
the parameters of the system and therefore can be set equal to zero,
and $\varepsilon_0\equiv\lambda^4/8\pi^2$. The quantities $A_i$ are
always positive and increase monotonically as $\tilde{m}_i^{2}$
increases. As we can see, the vacuum energy is determined by two
types of contributions. The contribution made by the parameters
$A_i$ (116) is due to the need of transition to the normal ordering
of operators in the free Hamiltonian $H_0$, and it exists also in
the linear theory, while another contribution to the vacuum energy
is determined by the nonlinearity of the system. The averaging of
the energy operator (113) over the vacuum state gives the vacuum
energy per unit volume:
\begin{equation} \label{117}
\begin{array}{cc}
\displaystyle{%
  \varepsilon_V\equiv\frac{E_V}{\Omega}=\varepsilon_0\sum_{i=1,2}\!   %
  \big[-\tilde{m}_i^4\ln\!\big(1+\tilde{m}_i^{-2}\big) +  %
  \ln\!\big(1+\tilde{m}_i^{2}\big) + \tilde{m}_i^{2}\big] + %
}\vspace{3mm}\\ %
\displaystyle{\hspace{0mm}%
  + \frac{g_0}{2}\bigg[\big(\chi_1^2+\chi_2^2\big)^2 - \frac{1}{2}\big(3\rho_1^2+3\rho_2^2+4\rho_{12}^2+2\rho_{1}\rho_{2}\big)\bigg].   %
}%
\end{array}
\end{equation}
Taking into account (97),\,(98), we find for $b$\,-phases
\begin{equation} \label{118}
\begin{array}{cc}
\displaystyle{%
  \frac{\varepsilon_V^{(b)}}{\varepsilon_0}= \frac{\tilde{\kappa}^2\big(\tilde{m}_1^2+\tilde{m}_2^2-2\tilde{\kappa}^2\big)}{2g_*} + %
  \ln\!\big(1+\tilde{m}_1^{2}\big)+\ln\!\big(1+\tilde{m}_2^{2}\big).   %
}%
\end{array}
\end{equation}
Here for  $b_l$\,-phase
$\tilde{m}_1^{2}=\tilde{m}_{l1}^{2},\,\tilde{m}_2^{2}=\tilde{m}_{l2}^{2}$,
and for $b_h$\,-phase
$\tilde{m}_1^{2}=\tilde{m}_{h1}^{2},\,\tilde{m}_2^{2}=\tilde{m}_{h2}^{2}$.
For $s$\,-phase, where
$\tilde{m}_1^{2}=\tilde{m}_2^{2}\equiv\tilde{m}_s^{2}$, the vacuum
energy density takes the form
\begin{equation} \label{119}
\begin{array}{cc}
\displaystyle{%
  \frac{\varepsilon_V^{(s)}}{\varepsilon_0}=2\bigg[\frac{\tilde{\kappa}^2\big(\tilde{m}_s^2-\tilde{\kappa}^2\big)}{2g_*} + %
  \ln\!\big(1+\tilde{m}_s^{2}\big)\bigg].   %
}%
\end{array}
\end{equation}
For $p$\,-phase, with taking account of (108) we have
\begin{equation} \label{120}
\begin{array}{cc}
\displaystyle{%
  \frac{\varepsilon_V^{(p)}}{\varepsilon_0}= \frac{\tilde{\kappa}^2\big(\tilde{m}_{p1}^2+\tilde{m}_{p2}^2-2\tilde{\kappa}^2\big)}{2g_*} - %
  \frac{\big(\tilde{m}_{p2}^2-\tilde{m}_{p1}^2\big)^2}{g_*} +
  \ln\!\big(1+\tilde{m}_{p1}^{2}\big)+\ln\!\big(1+\tilde{m}_{p2}^{2}\big).   %
}%
\end{array}
\end{equation}
The dependences of the vacuum energy density
$\tilde{\varepsilon}_V\equiv\varepsilon_V/\varepsilon_0$ on the mass
parameter $\tilde{\kappa}^2$ in different phases at $g_*=4$  are
shown in Fig.\,3. \newpage
\vspace{0mm} %
\begin{figure}[t!]
\vspace{0mm}  \hspace{0mm}
\includegraphics[width = 6.7cm]{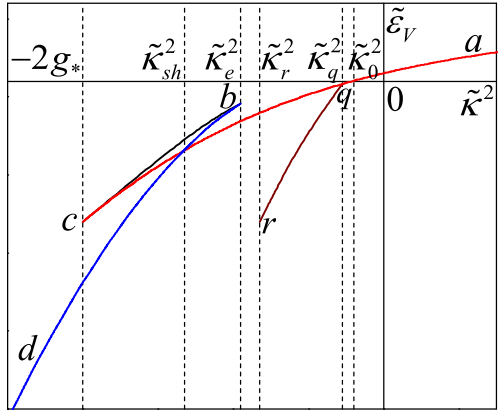} 
\vspace{-0mm} %
\caption{\label{fig03} 
The vacuum energy density
$\tilde{\varepsilon}_V\equiv\varepsilon_V/\varepsilon_0$ in different phases: %
$(ac)$ -- $s$\,-phase, $(bc)$ -- $b_l$\,-phase, $(bd)$ -- $b_h$\,-phase, $(qr)$ -- $p$\,-phase. %
Parameters at $g_*=4$: $\tilde{\kappa}_{sh}^2=-3.142$, $\tilde{\kappa}_{e}^2=-2.876$, %
$\tilde{\kappa}_r^2=-2.719$, $\tilde{\kappa}_q^2=-1.967$, $\tilde{\kappa}_0^2=-1.838$. %
}%
\end{figure}

In Fig.\,3 line $ac$ corresponds to the vacuum energy density of
$s$\,-phase, lines $bc$ and $bd$ to $b_l$\,-\, and $b_h$\,-phases,
and line $qr$ to $p$\,-phase. In some ranges of values of the
parameters $\tilde{\kappa}^2,\, g_*$ there may exist several phases.
In this case the phase with the lowest vacuum energy density is
realized. At $\tilde{\kappa}^2>0$ only  $s$\,-phase can exist,
moreover with a positive vacuum energy density. In the region
$\tilde{\kappa}_{e}^2 < \tilde{\kappa}^2 < 0$ $s$\,-phase also
exists, but at $\tilde{\kappa}_0^2 < 0$ its vacuum energy density
vanishes and at $\tilde{\kappa}^2 < \tilde{\kappa}_0^2$ it becomes
negative. Within the region $\tilde{\kappa}_{e}^2 < \tilde{\kappa}^2
< 0$ there is the region $\tilde{\kappa}_r^2 < \tilde{\kappa}^2 <
\tilde{\kappa}_q^2$ of existence of $p$\,-phase with the vacuum
energy less than that of $s$\,-phase.  In the interval
$\tilde{\kappa}_e^2 < \tilde{\kappa}^2 < \tilde{\kappa}_r^2$, %
if $g_* < 7.87$, again the existence of only $s$\,-phase is
possible. At $g_* > 7.87$ this region of $s$\,-phase disappears (see Fig.\,4). %

At $\tilde{\kappa}_e^2 < 0$ it becomes possible the existence of
also $b$\,-phases along with $s$\,-phase, however, in the interval
$\tilde{\kappa}_{sh}^2 < \tilde{\kappa}^2 < \tilde{\kappa}_e^2$, %
again $s$\,-phase is realized since here its vacuum energy is lower.
At point $\tilde{\kappa}_{sh}^2$ the vacuum energies of $s$\,-\, and
$b_h$\,-phases become equal, so that in the interval
$-2g_* < \tilde{\kappa}^2 < \tilde{\kappa}_{sh}^2$ %
it is also possible the existence of $s$\,-\, and $b$\,-phases, but
$b_h$\,-phase is realized. At $\tilde{\kappa}^2< -2g_*$ there exists
only $b_h$\,-phase, and the energy of its vacuum with a decrease in
the mass parameter $\tilde{\kappa}^2$ decreases according to the law
\begin{equation} \label{121}
\begin{array}{cc}
\displaystyle{%
  \frac{\varepsilon_V^{(h)}}{\varepsilon_0}\approx -2\frac{\tilde{\kappa}^4}{g_*} + \ln\!\big(1-2\tilde{\kappa}^2\big).   %
}%
\end{array}
\end{equation}
The range of parameters $\tilde{\kappa}^2,\, g_*$ where the
existence of $b_l$\,-phase would be energetically favorable is absent. %

The obtained formulas for the masses and vacuum energy density make
it possible to build a phase diagram of the states of the complex
scalar field on the plane $\big(\tilde{\kappa}^2,g_*\big)$, which is shown in Fig.\,4. %
\vspace{-1mm} %
\begin{figure}[h!]
\vspace{0mm}  \hspace{0mm}
\includegraphics[width = 6.6cm]{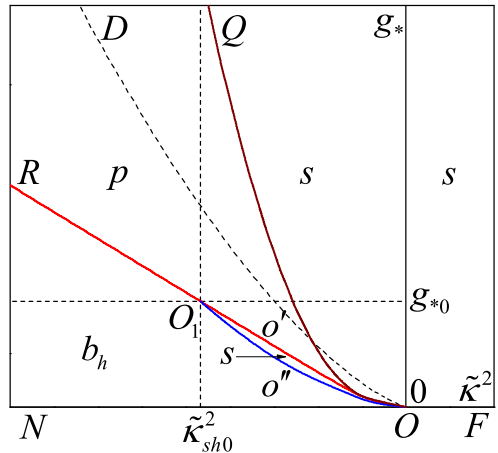} 
\vspace{-3mm} %
\caption{\label{fig04} 
Phase diagram of states of a complex scalar field on the plane $\big(\tilde{\kappa}^2,g_*\big)$. %
({\it 1}) To the right of curve $QOF$ and in the bounded area $O_1o''Oo'$ there is $s$\,-phase. %
({\it 2}) In the area $RO_1o'OQ$ there is $p$\,-phase. %
({\it 3}) In the area $RO_1o''ON$ there is $b_h$\,-phase. %
({\it 4}) On the curve $OD$ the vacuum energy density vanishes, on
the right it is positive and on the left it is negative. Coordinates
of point $O_1$: $(-4.67,7.87)$.
}%
\end{figure}

The question of the sign of the vacuum energy is of fundamental
importance for cosmology, because it determines the character of the
evolution of the Universe. In recent years, it has been discovered
that the expansion of the Universe proceeds with acceleration
[28,29]. As shown in this article on an example of a simple model,
the vacuum energy density is determined by the state of the field
itself and can be either positive or negative. Let us give a rough
estimate of the parameter $\varepsilon_0$, which is written in
ordinary units as $\varepsilon_0\equiv\hbar c\lambda^4\!\big/8\pi^2$, %
and $\lambda$ has the dimension of reciprocal length. As a cutoff
parameter, we take the inverse Compton wavelength of the $\pi_0$
meson $\lambda=m_\pi c/\hbar\approx 6.8\times 10^{12}\,{\rm cm}^{-1}$, %
where $m_\pi\approx 2.4\times 10^{-25}\,{\rm g}$. As a result, we
get the value $\varepsilon_0\approx 8\times 10^{32}\,{\rm erg}\!\cdot\!{\rm cm}^{-3}$. %
This is a very large value, which is many orders of magnitude
greater than the energy density in the Universe
$\varepsilon_{c0}\approx 8.5\times 10^{-9}\,{\rm erg}\!\cdot\!{\rm cm}^{-3}$ %
corresponding to the critical mass density
$\rho_{c0}\approx 10^{-29}\,{\rm g}\!\cdot\!{\rm cm}^{-3}$. %

However, as was shown  (117)\,--\,(120), the vacuum energy density
is determined by the quantity $\varepsilon_0$ multiplied by the
function, depending on the state of the field, which can be both
positive and negative or turn to zero for some parameters (curve
$OD$ in Fig.\,4). Of course, in order to compensate for the energy
difference of forty orders of magnitude, this field state function
must be close to zero with incredible accuracy. However, it should
be remembered that the field of only one nature was considered. If
the Universe is considered as a space filled with quantum nonlinear
interacting fields, both currently known and still unknown, then all
such fields will contribute to the vacuum energy density, and,
probably, with different signs. The estimates made allow to assume
that fields with broken symmetries make a negative contribution to
the energy the more significant, the heavier the particles
corresponding to the field. As a    result of the  compensation of
contributions of many fields, the vacuum energy density can have a
magnitude that differs greatly from the provided estimate. Thus, the
calculation of the energy of the physical vacuum state turns out to
be an incredibly difficult, if at all solvable from first principles
task. With the current state of the theory, the energy of the
physical vacuum state should be considered as a phenomenological
parameter.

\section{Interaction between particles. Perturbation theory}\vspace{-0mm} %
Let us consider in more detail the structure of the correlation
Hamiltonian (51), which describes the interaction between particles.
In the interaction representation it takes the form
\begin{equation} \label{122}
\begin{array}{l}
\displaystyle{%
   \hat{H}_C=\int\! d{\bf x}\bigg[\frac{g_0}{4}\big(\hat{\phi}_1^4+\hat{\phi}_2^4+2\hat{\phi}_1^2\hat{\phi}_2^2\big)+    %
   \frac{1}{2}\big(\kappa^2-B_1\big)\hat{\phi}_1^2 + \frac{1}{2}\big(\kappa^2-B_2\big)\hat{\phi}_2^2 -  %
   C_0\hat{\phi}_1\hat{\phi}_2 - p_1\hat{\phi}_1 - p_2\hat{\phi}_2 - V \bigg], %
}%
\end{array}
\end{equation}
where, as above, for brevity it is denoted $g_0= g\big/4!$. Through
the ``shifted'' operators (55), and taking into account (84) and
(86), the correlation Hamiltonian can be written as follows
\begin{equation} \label{123}
\begin{array}{ccccc}
\displaystyle{%
   \hat{H}_C=\frac{g_0}{4}\int\! d{\bf x}\Big\{ \big(\hat{\psi}_1^2+\hat{\psi}_2^2\big)^2  + %
             4\chi_1\hat{\psi}_1\big(\hat{\psi}_1^2+\hat{\psi}_2^2\big) + %
             4\chi_2\hat{\psi}_2\big(\hat{\psi}_1^2+\hat{\psi}_2^2\big) - %
}\vspace{2mm}\\ %
\displaystyle{%
  -2\big(3\tilde{\rho}_1+\tilde{\rho}_2\big)\hat{\psi}_1^2  %
  -2\big(3\tilde{\rho}_2+\tilde{\rho}_1\big)\hat{\psi}_2^2  %
  -8\tilde{\rho}_{12}\hat{\psi}_1\hat{\psi}_2 - %
}\vspace{2mm}\\ %
\displaystyle{%
  -4\Big[4\chi_1\big(\chi_1^2+\chi_2^2\big) + \chi_1\big(3\tilde{\rho}_1+\tilde{\rho}_2\big)+2\chi_2\tilde{\rho}_{12}\Big]\hat{\psi}_1 - %
}\vspace{2mm}\\ %
\displaystyle{%
  -4\Big[4\chi_2\big(\chi_1^2+\chi_2^2\big) + \chi_2\big(3\tilde{\rho}_2+\tilde{\rho}_1\big)+2\chi_1\tilde{\rho}_{12}\Big]\hat{\psi}_2 + %
}\vspace{2mm}\\ %
\displaystyle{%
  + 3\tilde{\rho}_1^2 +3\tilde{\rho}_2^2 + 2\tilde{\rho}_1\tilde{\rho}_2 + 4\tilde{\rho}_{12}^2 \Big\}. %
}%
\end{array}
\end{equation}
We introduce the definition of the normally ordered product of $n$
field operators by means of the relation
\begin{equation} \label{124}
\begin{array}{cc}
\displaystyle{%
  N\big(\hat{\psi}_1\hat{\psi}_2\ldots\hat{\psi}_n\big) =   %
  \hat{\psi}_1\hat{\psi}_2\ldots\hat{\psi}_n - \big\langle 0\big|\hat{\psi}_1\hat{\psi}_2\ldots\hat{\psi}_n\big|0\big\rangle.  %
}%
\end{array}
\end{equation}
Here the operators are taken in the interaction representation, and
$|0\rangle$ is the vacuum state vector of the system with the
Hamiltonian (57) that was defined above. This definition is
equivalent to the standard definition [21,22], according to which in
a normal product all creation operators stand to the left of
annihilation operators. Obviously, the definition (124) implies
\begin{equation} \label{125}
\begin{array}{cc}
\displaystyle{%
  \big\langle 0\big|N\big(\hat{\psi}_1\hat{\psi}_2\ldots\hat{\psi}_n\big)\big|0\big\rangle = 0. %
}%
\end{array}
\end{equation}
In addition, as follows from the form of the Hamiltonian (57), for odd $n$ there is always %
$\big\langle 0\big|\hat{\psi}_1\hat{\psi}_2\ldots\hat{\psi}_n\big|0\big\rangle = 0.$ %
In terms of the normal products and taking into account equations
(87),\,(88), the correlation Hamiltonian takes the form
\begin{equation} \label{126}
\begin{array}{ccccc}
\displaystyle{%
   \hat{H}_C=\frac{g_0}{4}\int\! d{\bf x}\,\Big\{  N\Big[\big(\hat{\psi}_1^2 +\hat{\psi}_2^2\big)^2\Big]  + %
             4\chi_1N\Big[\hat{\psi}_1\big(\hat{\psi}_1^2+\hat{\psi}_2^2\big)\Big] + %
             4\chi_2N\Big[\hat{\psi}_2\big(\hat{\psi}_1^2+\hat{\psi}_2^2\big)\Big] - %
}\vspace{2mm}\\ %
\displaystyle{%
  -2\big(3\tilde{\rho}_1+\tilde{\rho}_2\big)N\Big[\hat{\psi}_1^2\Big]  %
  -2\big(3\tilde{\rho}_2+\tilde{\rho}_1\big)N\Big[\hat{\psi}_2^2\Big]  %
  -8\tilde{\rho}_{12}N\Big[\hat{\psi}_1\hat{\psi}_2\Big] + %
}\vspace{2mm}\\ %
\displaystyle{%
  +\,4\chi_1\big[\big(\chi_1^2+\chi_2^2\big)+\kappa^2g_0^{-1}\big]N\Big[\hat{\psi}_1\Big] %
  +\,4\chi_2\big[\big(\chi_1^2+\chi_2^2\big)+\kappa^2g_0^{-1}\big]N\Big[\hat{\psi}_2\Big] \Big\}. %
}%
\end{array}
\end{equation}
Since in the Goldstone approximation
$\chi_1^2+\chi_2^2=-\kappa^2g_0^{-1}$ and
$\tilde{\rho}_1=\tilde{\rho}_2=\tilde{\rho}_{12}=0$, then (126) in this case %
\begin{equation} \label{127}
\begin{array}{ccccc}
\displaystyle{%
   \hat{H}_C=\frac{g_0}{4}\int\! d{\bf x}\,\Big\{  N\Big[\big(\hat{\psi}_1^2 +\hat{\psi}_2^2\big)^2\Big]  + %
             4\chi_1N\Big[\hat{\psi}_1\big(\hat{\psi}_1^2+\hat{\psi}_2^2\big)\Big] + %
             4\chi_2N\Big[\hat{\psi}_2\big(\hat{\psi}_1^2+\hat{\psi}_2^2\big)\Big] \Big\}. %
}%
\end{array}
\end{equation}
The interaction Hamiltonian (127) has the same form as for the
classical field, except that it is written in terms of the normally
ordered products. In terms of the fields $\hat{u}_1,\,\hat{u}_2$,
describing particles with certain masses, the correlation
Hamiltonian takes the form
\begin{equation} \label{128}
\begin{array}{ccc}
\displaystyle{%
   \hat{H}_C=\frac{g_0}{4}\int\! d{\bf x}\,\Big\{  N\Big[\big(\hat{u}_1^2 +\hat{u}_2^2\big)^2\Big]  + %
             4a_-N\Big[\hat{u}_1\big(\hat{u}_1^2+\hat{u}_2^2\big)\Big] + %
             4a_+N\Big[\hat{u}_2\big(\hat{u}_1^2+\hat{u}_2^2\big)\Big] + %
}\vspace{2mm}\\ %
\displaystyle{%
  +\bigg(A-2\,\frac{m_1^2}{m_2^2}\,a_+^2\bigg)N\Big[\hat{u}_1^2\Big]  %
  +\bigg(A-2\,\frac{m_1^2}{m_2^2}\,a_-^2\bigg)N\Big[\hat{u}_2^2\Big]  %
  +4\frac{m_1^2}{m_2^2}\,a_+a_-N\Big[\hat{u}_1\hat{u}_2\Big] + %
}\vspace{2mm}\\ %
\displaystyle{%
  +\,4Aa_-N\big[\hat{u}_1\big] +\,4Aa_+N\big[\hat{u}_2\big] \Big\}, %
}%
\end{array}
\end{equation}
where the notation is introduced
\begin{equation} \label{129}
\begin{array}{ccccc}
\displaystyle{%
   A\equiv \frac{1}{2g_0}\big(2\kappa^2+m_2^2\big), \qquad %
   a_+\equiv\alpha\chi_2 +\beta\chi_1, \qquad a_-\equiv\alpha\chi_1 -\beta\chi_2. %
}%
\end{array}
\end{equation}
The parameters $\alpha,\beta$ are defined above by relations (62).
In the Goldstone approximation $A=0$ and $m_1^2=0$, so that we again
arrive at an expression similar to (127)
\begin{equation} \label{130}
\begin{array}{ccccc}
\displaystyle{%
   \hat{H}_C=\frac{g_0}{4}\int\! d{\bf x}\,\Big\{  N\Big[\big(\hat{u}_1^2 +\hat{u}_2^2\big)^2\Big]  + %
             4a_-N\Big[\hat{u}_1\big(\hat{u}_1^2+\hat{u}_2^2\big)\Big] + %
             4a_+N\Big[\hat{u}_2\big(\hat{u}_1^2+\hat{u}_2^2\big)\Big] \Big\}, %
}%
\end{array}
\end{equation}
and here $a_-=0,\, a_+=\sqrt{\chi_1^2+\chi_2^2}$\,.

As we can see, the interaction Hamiltonian has the form of the sum
of two terms, one of which contains terms of the third order, and
the other of the fourth order in the field operators
$\hat{H}_C=\int\! d{\bf x}\big[\hat{H}_C^{(4)}(x)+\hat{H}_C^{(3)}(x)\big]$. %
The averages of the introduced normal products over the vacuum state
are equal to zero, so that the condition
$\big\langle 0\big|\hat{H}_C\big|0\big\rangle = 0$ %
is automatically satisfied. Thus, in this approach the normal form
of the interaction Hamiltonian is not postulated initially, but
arises as a consequence of the choice of the self-consistent field
model as the main approximation, in which the effects of zero
fluctuations are already taken into account.

The $n$-th order contribution to the $S$\,-\,matrix
$S=\sum_{n=0}^\infty S^{(n)}$ can be represented as
\begin{equation} \label{131}
\begin{array}{ccc}
\displaystyle{%
  S^{(n)}=\frac{(-i)^n}{n!}\int\! dx_1\ldots dx_n\,
  T\big[ \hat{H}_C^{(4)}(x_1)\ldots \hat{H}_C^{(4)}(x_n) + %
         \hat{H}_C^{(3)}(x_1)\ldots \hat{H}_C^{(3)}(x_n)\, + %
}\vspace{2mm}\\ %
\displaystyle{%
  + \sum_{m=1}^{n-1}C_n^m \hat{H}_C^{(4)}(x_1)\ldots \hat{H}_C^{(4)}(x_m)\hat{H}_C^{(3)}(x_{m+1})\ldots \hat{H}_C^{(3)}(x_n), %
}%
\end{array}
\end{equation}
where
$\displaystyle{\hat{H}_C^{(4)}(x)}=\frac{g_0}{4}N\Big[\big(\hat{u}_1^2 +\hat{u}_2^2\big)^2\Big]$,\, %
$\displaystyle{\hat{H}_C^{(3)}(x)}=g_0a_-N\Big[\hat{u}_1\big(\hat{u}_1^2
+\hat{u}_2^2\big)\Big]+g_0a_+N\Big[\hat{u}_2\big(\hat{u}_1^2 +\hat{u}_2^2\big)\Big]$,\, %
$C_n^m$ are the binomial coefficients, $T$ is the chronological
operator. The perturbation theory and diagram technique are
constructed in a standard way, similar to how it was done for a real
scalar field [19] and interacting Fermi and Bose fields [20].

\newpage%
\section{Discussion. Conclusions}\vspace{-0mm} %
In modern quantum field theory and elementary particle physics, the
states of fields with broken symmetries play a fundamental role. To
describe such states it is necessary to consistently and
systematically take into account the nonlinearity of the fields
without the assumption of a weak nonlinearity. Meanwhile, the modern
description of quantum fields is based on the model of free
noninteracting fields, and the nonlinearity due to the interaction
of fields is taken into account by perturbation theory [21,22]. Such
a decomposition of the Lagrangian of the system into the main
approximation and perturbation is equivalent to choosing the ideal
gas model as the main approximation in the theory of nonrelativistic
many-particle systems. But the ideal gas model does not allow to
correctly describe the phase transitions that are caused by the
interaction between particles. Even a thermodynamically correct
description of the phenomenon of Bose-Einstein condensation, which
was discovered in the framework of the ideal gas model, requires
accounting for the interparticle interaction [30].

Therefore, to construct a consistent perturbation theory in the
presence of symmetry breaking, it is natural to choose as the main
approximation such a model that would allow the existence of states
with different symmetries and at the same time retain the
possibility of describing the field in the language of particles or,
more precisely, quasiparticles. As shown in this work on an example
of a complex scalar field, the approximation that satisfies the
above requirements is the relativistic model of a self-consistent
field. An important feature of this model is that when the
self-consistent field is chosen as the main approximation, the
particle's interaction Hamiltonian automatically acquires a normally
ordered form without involving any additional approximations. This
makes it possible to construct a perturbation theory by means of
standard methods and, at the same time, systematically take into
account the effects caused by zero field oscillations, which
undoubtedly play a fundamental role. In addition, with account of
nonlinear effects there can be calculated the vacuum energy density,
which proves to depend on the state of the field itself.

Let us summarize the main results of the work. The relativistic
self-consistent field model for the complex scalar field is
formulated and the perturbation theory is constructed on its basis.
It is shown that the account of the pair anomalous averages,
violating the phase symmetry, leads to the appearance of masses for
particles, so that there are no massless bosons in the theory. The
particle masses are calculated both in the symmetric state and in
the states with broken phase symmetry. The vacuum energy density in
different phases is calculated and it is shown that it can be both
positive and negative. The phase diagram is constructed in the
coordinates mass parameter -- interaction constant.

The author is grateful to A.A.\,Soroka for the help in numerical
calculations and article preparing.



\end{document}